\documentclass{aa}  

\usepackage{graphicx}
\usepackage{siunitx}
\usepackage{caption} 
\usepackage{subcaption}    
\usepackage{txfonts}
\makeatletter
\newcommand\footnoteref[1]{\protected@xdef\@thefnmark{\ref{#1}}\@footnotemark}
\makeatother
\newcommand{\ltt}{LTT-9779}
\newcommand{\lttb}{LTT-9779\,b}
\usepackage{hyperref}
\usepackage{mathtools}
\usepackage{multirow}
\titlerunning{No TiO detected in \lttb's high resolution reflection spectrum}

\begin{document}

   \title{No TiO detected in the hot Neptune-desert planet LTT-9779 b in reflected light at high spectral resolution}

   \subtitle{}

   \author{Sophia R. Vaughan\inst{1,2}
          \and
          Jayne L. Birkby\inst{2}
          \and
          Natasha E. Batalha\inst{3}
          \and
          Luke T. Parker\inst{2}
          \and
          Haochuan Yu\inst{2}
          \and
          Julia V. Seidel\inst{4,5}
          \and
          Michael Radica\inst{6}
          \and
          Jake Taylor\inst{2}
          \and
          Laura Kreidberg\inst{1}
          \and
          Vivien Parmentier\inst{2,5}
          \and
          Sergio Hoyer\inst{7}
          \and
          James S. Jenkins\inst{8,9}
          \and
          Annabella Meech\inst{2,10}
          \and
          Ricardo Ram\'irez Reyes\inst{11}
          \and
          Lennart van Sluijs\inst{12}
          }

   \institute{Max-Planck-Institut f\"ur Astronomie, K\"onigstuhl 17, 69117 Heidelberg, Germany
         \and
             Department of Physics, University of Oxford, Oxford, OX1 3RH, UK
         \and 
             NASA Ames Research Center, Moffett Field, CA 94035, USA
         \and
             European Southern Observatory, Alonso de Córdova 3107, Vitacura, Región Metropolitana, Chile 
         \and
             Laboratoire Lagrange, Observatoire de la Côte d’Azur, CNRS, Université Côte d’Azur, Nice, France
         \and
             Department of Astronomy \& Astrophysics, University of Chicago, 5640 South Ellis Avenue, Chicago, IL 60637, USA
         \and
             Aix Marseille Univ, CNRS, CNES, LAM, Marseille, France
         \and
             Instituto de Estudios Astrof\'isicos, Facultad de Ingenier\'ia y Ciencias, Universidad Diego Portales, Av. Ej\'ecito 441, Santiago, Chile
         \and
             Centro de Astrof\'isica y Tecnolog\'ias Afines (CATA), Casilla 36-D, Santiago, Chile
         \and
             Center for Astrophysics, Harvard \& Smithsonian, 60 Garden St, Cambridge, MA 02138, USA
         \and
             Departamento de Astronom\'ia, Universidad de Chile, Camino el Observatorio 1515, Las Condes, Santiago, Chile
         \and
             Department of Astronomy, University of Michigan, 1085 South University Avenue, Ann Arbor, MI 48109, USA
             }

   \date{Received 15th Sep 2025; accepted 7th Nov 2025}
 
  \abstract
   {LTT-9779\,b is an inhabitant of the hot Neptune desert and one of only a few planets with a measured high albedo. Characterising the atmosphere of this world is the key to understanding what processes dominate in reducing the number of short-period intermediate mass planets, creating the hot Neptune desert. We aim to characterise the reflected light of LTT-9779\,b at high spectral resolution to break the degeneracy between clouds and atmospheric metallicity. This is key to interpreting its mass loss history which may illuminate how it kept its place in the desert. We use the high resolution cross-correlation spectroscopy technique on four half-nights of ESPRESSO observations in 4-UT mode (16.4-m effective mirror) to constrain the reflected light spectrum of \lttb. We do not detect the reflected light spectrum of \lttb\ despite these data having the expected sensitivity at the level 100 ppm. Injection tests on the post-eclipse data indicate that TiO should have been detected for a range of different equilibrium chemistry models. Therefore this non-detection suggests TiO depletion in the western hemisphere however, this conclusion is sensitive to temperature which impacts the chemistry in the upper atmosphere and the reliability of the line list. Additionally, we are able to constrain the top of the western cloud deck to $P_{\text{top, western}}<10^{-2.0}$ bar and the top of the eastern cloud deck $P_{\text{top, eastern}}<10^{-0.5}$ bar, which is consistent with the predicted altitude of MgSiO$_3$ and  Mg$_2$SiO$_4$ clouds from JWST NIRISS/SOSS. While we do not detect the reflected light spectrum of \lttb, we have verified that this technique can be used in practice to characterise the high spectral resolution reflected light of exoplanets so long as their spectra contain a sufficient number of deep spectral lines. Therefore this technique may become an important cornerstone of exoplanet characterisation with the ELT and beyond.}
   {}{}{}{}


   \maketitle
%

\section{Introduction}
\label{sec:introduction}

\lttb\ is a rare example of a mature (Age $\sim2$ Gyr), intermediate-mass ($M\sim29 M_\oplus$), ultra-short period ($P\sim0.79$~d) planet with an atmosphere that currently comprises approximately $9\%$ of the planet's mass \citep[][see Table \ref{tab:lttsystem} for more details]{Jenkins2020}. Despite this system's age, it is possible that is atmosphere is still undergoing mass loss. There is an observed dearth of planets such as \lttb\ \citep[e.g.][]{Mazeh2016} which is often referred to as the `hot Neptune desert'. The root cause of the hot Neptune desert is uncertain and it could be the result of one or more underlying processes such as photo-evaporative atmospheric loss \citep{Lopez2013, Owen2013, Owen2016, Owen2017}; core powered mass loss of the atmospheres \citep{Ginzburg2018, Gupta2019, Gupta2020, Gupta2021, Gupta2022}; planet migration leading to tidal disruption \citep{Matsakos2016, Owen2018}; and biases in the outcomes of planet formation \citep{Lee2021, Lee2022, Nielsen2025}. Additionally, since \lttb\ lies close to the inner edge of the hot Neptune desert it may be now, or in the past, subject to Roche Lobe Overflow \citep{Jackson2017, Koskinen2022}. If any or a combination of these process are responsible for shaping the desert, then some quirk of \lttb's atmosphere, space environment, system age and/or history must be responsible for its current place there. By studying this system, we learn more about which processes allow its atmosphere to survive and how they shape the hot Neptune desert.

\begin{table}
	\begin{center}
    \caption{Summary of the \ltt\ system.}
    \label{tab:lttsystem}
	\begin{tabular}{lrc}
		\hline
		\ltt & Value & Ref. \\
        \hline
        \hline
        Mass ($M_\odot$) & $1.02^{+0.02}_{-0.02}$ & 1 \\
        Temperature (K) & $5443^{+14}_{-13}$ & 1 \\
        Age (Gyr) & $2.0^{+1.3}_{-0.9}$ & 1\\
        Stellar Type & G7 V & 1 \\
        Magnitude V band & 9.79 & 2 \\
        $v \sin(i)$ (km s$^{-1}$)&  $1.81^{+0.06}_{-0.07}$ & 1 \\
        Rotation Period (day) & 45 & 1 \\
		\hline
		\lttb & Value & Ref. \\
        \hline
        \hline
        Orbital Period (days) & $0.79206410\pm0.00000014$& 3 \\
        Semi-major axis (au) & $0.01679^{+0.00014}_{-0.00012}$ & 1\\
        Mass ($M_{\text{jup}}$) & $0.09225^{+0.00245}_{-0.00255}$ & 1\\
        Mass ($M_\oplus$) & $29.32^{+0.78}_{-0.81}$ & 1 \\
        Radius ($R_{\text{jup}}$) & $0.421\pm0.021$ & 1\\
        Radius ($R_\oplus$) & $4.72\pm0.23$ & 1\\
        Effective Temperature (K) & $1978\pm19$ & 1\\
		\hline 
	\end{tabular}
	\end{center}
        \textbf{References.} (1) \citet{Jenkins2020}; (2) \citet{Hog2000}; (3) \citet{Edwards2023a}
\end{table}

This system has already been the focus of numerous studies which have shown it to have a number of unusual properties. \citet{Fernandez2023} used XMM-Newton observations to show that the XUV output of \ltt\ was much lower than is typical. This would allow \lttb\ to retain an atmosphere to this day if it started with an envelope fraction greater than 10\%, although it would still be undergoing photo-evaporation. \citet{Edwards2023a}, \citet{Vissapragada2024} and \cite{Radica2024a} did not detect signatures of mass loss in the $\SI{1.083}{\micro\meter}$ Helium line using transit observations taken with WFC3/Hubble, WINERED/Magellan II and NIRISS/JWST respectively. This either limits the mass loss rate to below the upper limit predicted by \citet{Fernandez2023} or implies that this line is not excited. Alternatively, \citet{Radica2024a} posited that the clouds may be suppressing atmospheric escape by lowering the temperature of the atmosphere. To determine if \lttb\ is undergoing mass loss and at what rate, more information on its atmosphere is needed. 

Several studies have measured the atmospheric metallicity of \lttb\ \citep{Crossfield2020, Hoyer2023, Edwards2023a, Radica2024a, Coulombe2025, Reyes2025, Zhou2025, Ashtari2025} but the constraints are not strong and the results are not fully consistent with each other. For example, \citet{Coulombe2025} using JWST NIRISS/SOSS reflected light phase curves suggested the atmosphere has a metallicity less than $30\times$ solar while \citet{Crossfield2020} using a $4.5\mu m$ Spitzer phase curve suggested instead that the metallicity was greater than $30\times$ solar. The works of \citet{Hoyer2023}, \citet{Radica2024a}, \citet{Reyes2025} and \citet{Ashtari2025} all favour high metallicities with the former requiring $>400\times$ solar metallicity to explain the optical albedo, the next two requiring metallicities $20-850\times$ and $>180\times$ solar respectively to explain the flat transmission spectrum and the last suggesting $>500\times$ solar based on the JWST NIRSpec G395H phase curve. In contrast, \citet{Edwards2023a} presented WFC3/Hubble transit observations which suggested less than $0.1\times$ solar metallicity however, these data were reanalysed by \citet{Zhou2025} whom found similar values until they included the JWST NIRISS/SOSS data from \citet{Radica2024a} which raised the metallicity to $275\times$ solar. The lack of strong, consistent constraints on \lttb's metallicity is a strong barrier against understanding the atmosphere of this world and therefore its place in the desert due to its link with the planet's formation history and mass loss \citep[e.g.][]{Fortney2013, Lee2016, Owen2017, Louca2025}. 

In additional to the metallicity, several studies also aimed to investigate the abundances of individual atmospheric species. \citet{Dragomir2020} showed Spitzer secondary eclipse data strongly favoured models containing CO or CO$_2$ while \citet{Coulombe2025} used JWST NIRISS/SOSS phase curves to constrain the water abundance. Both \citet{Edwards2023a} and \citet{Zhou2025} studied the same WFC3/Hubble transits but were not able to robustly constrain the presence of atmospheric species and got inconsistent results. Further to this, while hot-Jupiters of similar equilibrium temperature to \lttb\ tend to have inversions \citep{Parmentier2016}, Spitzer and JWST observations indicate that the day-side atmosphere has a non-inverted temperature-profile \citep{Dragomir2020, Coulombe2025}. This can imply a lack of absorbers such as VO and TiO at high altitudes. However this is an indirect measurement so additional data would be needed to confirm the lack of such species in the upper atmosphere of \lttb. 

The difficulty in constraining the metallicity and detecting the spectral features of single species is likely due to the presence of clouds in the atmosphere. These essentially hide the lower layers of the atmosphere and can mute the size of the spectral features making them harder to detect and constrain. The altitude of the cloud deck determines how much of the atmosphere the clouds obscure and so how much the spectral features are muted. It is, however, difficult to constrain altitude of the cloud deck since spectral feature can also be muted if metallicity is decreased leading to degeneracies between these two parameters.

\lttb\ has one more unusual trait, it is one of only a handful of known highly reflective worlds \citep{Hoyer2023, Coulombe2025, Saha2025}. JWST NIRISS/SOSS phase curves have shown this reflection to vary spatially with the western limb having a geometric albedo ($A_g$) of $0.8$ and the eastern limb $A_g=0.4$ \citep{Coulombe2025}. \citet{Coulombe2025} showed their data was consistent with Mg$_2$SiO$_4$/MgSiO$_3$ clouds that are primarily present on the western hemisphere of the planet. This spatial distribution is consistent with common trends seen in hot-Jupiters where the cooler temperatures on the evening limb cause cloud formation \citep{Fu2025}. Further to this, the non-detection of the planet's eclipse with UVES also suggests Mg$_2$SiO$_4$/MgSiO$_3$ clouds \citep{Radica2025}. However, the altitude of these clouds remains unconstrained \citep{Radica2025, Coulombe2025}.

\begin{figure}
	\includegraphics[width=\columnwidth]{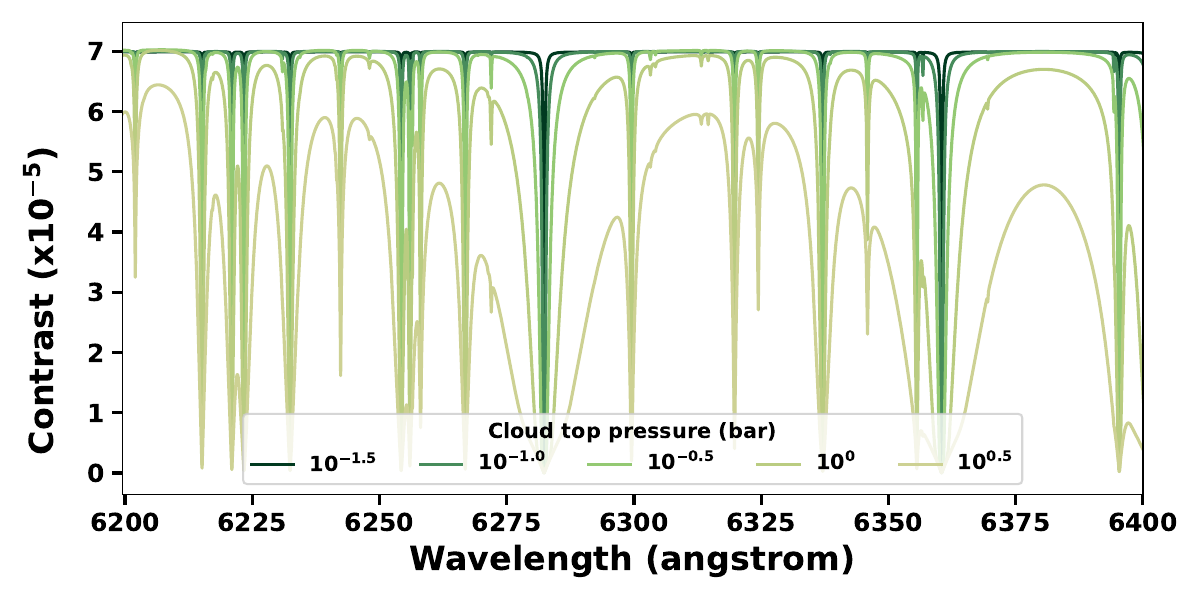}
    \caption{Spectrum of a hydrogen-helium atmosphere with a uniform volume mixing ratio of Fe of 0.0001 (see Section \ref{sec:searchgridtemplates}). Each line represents a model with a grey cloud deck at a different altitude which is represented by the cloud top pressure.}
    \label{fig:linewidth}
\end{figure}

The reflected light of an exoplanet is comprised of the light of its host star which has been scattered by the exoplanet and its atmosphere into the direction of the observer. Thus the reflected stellar light becomes imprinted with absorption features from the exoplanet's dayside atmosphere. If the exoplanet has a highly reflective cloud deck, the spectral lines are formed primarily by the atmosphere above the clouds. The clouds themselves do not produce spectral lines in this wavelength range, rather they change the continuum of the spectrum. This means the depth of the spectral features, assuming the reflectivity of the clouds is fixed, is dependent both on the abundance of a given species in the upper atmosphere and the altitude of the cloud deck. Current data has not had the sensitivity to break the degeneracy between metallicity and cloud deck altitude. However, at high spectral resolution, this degeneracy is broken because the altitude of the cloud deck alters the shape of the spectral lines. \citet{Gandhi2020} shows examples of the breaking of this degeneracy for transmission spectra and Figure \ref{fig:linewidth} shows the same for reflection spectra using models from Section \ref{sec:searchgridtemplates}.

High resolution cross-correlation spectroscopy \citep[HRCCS][]{Snellen2010} is a well established technique for obtaining and characterising high-resolution exoplanet spectra which has been predominantly used on the transmission and emission spectra of exoplanets \citep[see][]{Birkby2018, Snellen2025}. Attempts have been made to study reflection spectra with this technique \citep{Martins2015, Hoeijmakers2018a, Scandariato2021, Spring2022} however no robust detections were obtained. \citet{Martins2015} claimed a detection of the reflected light of 51 Pegasi b, but \citet{Scandariato2021} and \citet{Spring2022} later showed this to be an unfortunate false positive. The lack of successful detection of the reflected light of an exoplanet at high spectral resolution has so far been attributed to the low albedos of the previously targeted exoplanets. However, \lttb\ is known to have a high albedo \citep{Hoyer2023, Coulombe2025, Saha2025} making it a good candidate for this technique. 

Beyond learning more about the atmosphere and clouds of \lttb, demonstrating the HRCCS technique in reflected light on \lttb\ is of wider importance to the future of exoplanet characterisation. Earth-like planets have low transit probabilities and are relatively cool with little thermal emission in optical and near infrared. Therefore the reflected light of these worlds will form an important part of the characterisation of their atmospheres. For the extremely large telescope (ELT), this characterisation will largely focus on the use of high resolution techniques such as HRCCS and the related molecule mapping technique \citep{Snellen2015, Hoeijmakers2018b}.

In this work, we aim to demonstrate the HRCCS technique on the reflected light of \lttb. In Section \ref{sec:observations} we discuss the observations and in Section \ref{sec:orbit} we show a refitting of the planet's orbit using these data. In Section \ref{sec:datareduction}, we discuss the HRCCS data reduction and the analysis of the reduced data is discussed in Section \ref{sec:recoverplanet}. In Section \ref{sec:results} we discuss our results. Finally, we discuss the atmospheric constraints imposed by this data and highlight future directions in Section \ref{sec:discussion} and conclude in Section \ref{sec:conclusions}.

\section{Observations}
\label{sec:observations}

The HRCCS technique requires a time-series of high resolution, high signal-to-noise spectra of the combined star-planet system so that the planet's changing Doppler shift can be used to isolate its spectrum. We observed the planet host star \ltt\ (G7V, $V=9.76$ mag) for four half-nights, each covering before, during\footnote{While the planet is not visible during secondary eclipse, these phases provide a measure of the star-only spectrum for comparison.}, and after secondary eclipse of \lttb, using the echelle spectrograph for rocky exoplanets and stable spectroscopic observations \citep[ESPRESSO;][]{Pepe2021} on the very large telescope (VLT) at Paranal Observatory, Chile (PI: Vaughan; Program ID: 112.25T7). This phase coverage is possible due to the short approximately $19$ hour orbital period which allows around $25\%$ of the orbit to be observed in each half-night. 

These observations use ESPRESSO's `4-UT mode'. In this mode, the ESPRESSO instrument is fed light from all four of the VLT's `unit telescopes' (UT) simultaneously. The light reaches the instrument in a bundle of four fibres and the combined light is projected onto the spectrograph. ESPRESSO was designed to operate in the slit-limited regime of resolution so the wider fibre bundle results in a lower spectral resolution compared with the available 1-UT modes. However, the signal-to-noise of the resulting spectrum is significantly higher primarily due to the larger collecting area\footnote{Using all four UT's is equivalent to using a primary mirror $\SI{16.4}{\meter}$ in diameter.} provided by using multiple telescopes. Overall, the signal-to-noise of a spectrum increases by a factor of approximately $2.7$ when 4-UT mode is used over 1-UT mode. This observing mode has been used for exoplanet transit characterisation in the past \citep{Borsa2021, Seidel2025, Prinoth2025} but not to study an exoplanet in reflected light. As part of the observing proposal, we used a modified version of the ESPRESSO exposure time calculator to simulate these observations at the appropriate spectral resolution. These simulated observations contained the tellurics, seeing and airmass trends as well as the appropriate read, dark, and photon noise. \lttb's spectrum was modelled assuming the planet was in chemical equilibrium and so it contained a large number of deep TiO lines. When the simulated observations were analysed, using the same techniques as would be applied to real data, the signal of the planet's spectrum was recovered at a signal-to-noise of approximately ten.

In 4-UT mode, ESPRESSO covers a simultaneous wavelength range of $\SI{380}{\nano\meter}$ to $\SI{788}{\nano\meter}$ at a spectral resolution of $R\sim70,000$ \citep{Pepe2021}. Therefore the instrumental broadening (or resolution element) is approximately $\SI{4.3}{\kilo\meter\per\second}$ and the atmospheric absorption lines in the planet’s spectrum\footnote{See Section \ref{sec:spectrageneration} for a discussion on the broadening on the reflected stellar lines.} are dominated by this instrumental broadening since the planetary spin, assuming tidal locking, results in a broadening of only around $\SI{2.7}{\kilo\meter\per\second}$. We used an integration time of $\SI{300}{\second}$ during the first night of observations due to the reduced seeing conditions which leads to less light entering the fibre. This integration time leads to a Doppler smearing of approximately $\SI{5.3}{\kilo\meter\per\second}$ for phases outside of secondary eclipse. This is slightly larger than the resolution element, however we elected to favour obtaining a higher signal-to-noise in the continuum. Subsequent nights experienced better seeing so an integration time of $\SI{200}{\second}$ was used instead to avoid saturation. The Doppler smearing in these observations is less than $\SI{4.3}{\kilo\meter\per\second}$ for the observed fraction of the orbit. The third night of observation had to be halted due to precipitation at the VLT site and was not restarted. A summary of the observations is given in Table \ref{tab:observations}.

\begin{table*}
	\begin{center}
	\caption{Summary of the \ltt\ eclipse observations.}
	\label{tab:observations}
	\begin{tabular}{lccccccc} 
		\hline
		   Date & Integration Time & \multicolumn{3}{c}{Frames} & \multicolumn{2}{c}{Orbital phase} & S/N of Order 47 $\dagger$\\
           & (s) & Total & Pre-eclipse & Post-eclipse & Start & End & \\
		\hline
        \hline
		10th Oct 2023 & 300 & 64 & 28 & 26 & 0.324 & 0.650 & 171\\
		14th Oct 2023 & 200 & 89 & 32 & 44 & 0.372 & 0.673 & 170\\
		21st Oct 2023 & 200 & 43 & 43 & 0 & 0.319 & 0.463 & 153\\
		17th Nov 2023 & 200 & 76 & 43 & 19 & 0.337 & 0.592 & 153\\
		\hline
	\end{tabular}
    \end{center}
    $\dagger$ Order 47 has a central wavelength $\lambda_{\text{central}} = \SI{527.48}{\nano\meter}$ and contains the peak of \ltt's blackbody spectrum.
\end{table*}

The spectra are extracted from the raw data using ESOReflex v2.11.5 and ESPRESSO pipeline v3.1.0. This pipeline performs a number of standard calibrations such as bias correction, flat fielding, wavelength calibration and background subtraction. It extracts the spectra in the barycentric rest frame and into either a single combined spectrum or into their individual spectral orders. We choose to use the latter to reduce the amount of preprocessing which can alter the systematics in the data. In addition to extracting the spectra the ESPRESSO pipeline also computes the cross-correlation function of the host star for each observation using a binary mask (G9) selected based on the \ltt's stellar type (G7V). This is then used to calculate a precise stellar radial velocity for each observation.

\section{Refitting the orbit of \lttb}
\label{sec:orbit}

HRCCS usually requires the cross-correlation of each exposure to be stacked in the planetary rest frame so that the signal of the planet's spectrum can be confidently detected. This requires an accurate planetary ephemeris to predict the orbital phase, and thus the velocity of the planet. The error in the orbital phase prediction comes primarily from the uncertainty in the orbital period, which can result in a large discrepancy between predictions and the true orbital phase if the planet has completed many orbits since the measurement. There are three ephemerides for \lttb; the discovery ephemeris from \citet{Jenkins2020} and two updates \citep{Edwards2023a, Kokori2023}. The \citet{Jenkins2020} ephemeris has a $\SI{0.8}{\second}$ error on the orbital period which propagates to an uncertainty in the orbital phase of $0.0054$ after one Earth year or $\sim 0.028$ during the epochs of our ESPRESSO observations. It is thus unsuitable for our analysis. The other two ephemerides \citep{Edwards2023a, Kokori2023} have errors on the orbital period of $\SI{0.012}{\second}$ and $\SI{0.024}{\second}$ respectively and are also more recent, meaning less orbits have occurred since they were measured. Therefore the uncertainty in the orbital phase for both ephemerides is less than $0.001$ when propagated to our observation epochs and the predictions are within the one sigma errors of each other. This corresponds to an approximate error in the planet's velocity of approximately $\SI{0.7}{\kilo\meter\per\second}$ at secondary eclipse. At $R=70000$, the velocity resolution is $\SI{4.3}{\kilo\meter\per\second}$ or approximately six times greater than the error which is sufficiently accurate for our analysis. 

However, with the precise stellar radial velocities measured by the ESOReflex pipeline and high temporal cadence of these observations, it is possible to constrain parameters such as the eccentricity of \lttb's orbit. A non-zero eccentricity could be an indication of high-eccentricity migration \citep{Matsakos2016, Owen2018}, one of the processes that might form the hot Neptune desert. A small eccentricity can also raise the internal temperature via tidal heating \citep{Leconte2010} which would affect the interpretation of atmospheric parameters.

Just using the radial velocities from these observations does not constrain the planet's orbit due to the limited phase coverage. To create a more complete radial velocity curve, the radial velocities from three nights of 1-UT ESPRESSO transit observations (PI: Jenkins, Program ID:103.2028 and PI: Ramirez Reyes, Program ID:108.22FQ) as well as the tabulated radial velocities from HARPS\footnote{The CORALIE values presented in \citet{Jenkins2020} are not included as these have much larger errors.} presented in \citet{Jenkins2020} are included. Additionally, the short cadence ($<2$mins) TESS light curves, extracted with \texttt{lightkurve} python package \citep{Lightkurve2018}, are fit simultaneously with these data.

The orbit is modelled using the \texttt{exoplanet} python package \citep{Foreman-Mackey2021} and to compute the best fit we use the Bayesian inference package \texttt{pymc3} \citep{Salvatier2016}. We select the \texttt{pymc3} NUTS sampler to evaluate the posteriors of each fit which we use to calculate the uncertainties. We choose not to use nested sampling as the results can be less robust in this instance. We preform two fits, one with the eccentricity fixed to zero and a second where eccentricity is a free parameter, the results of which are shown in Table \ref{tab:orbitfit}. Both orbital fits assume \lttb\ is the only planet in the system. 

\begin{figure}
	\includegraphics[width=\columnwidth]{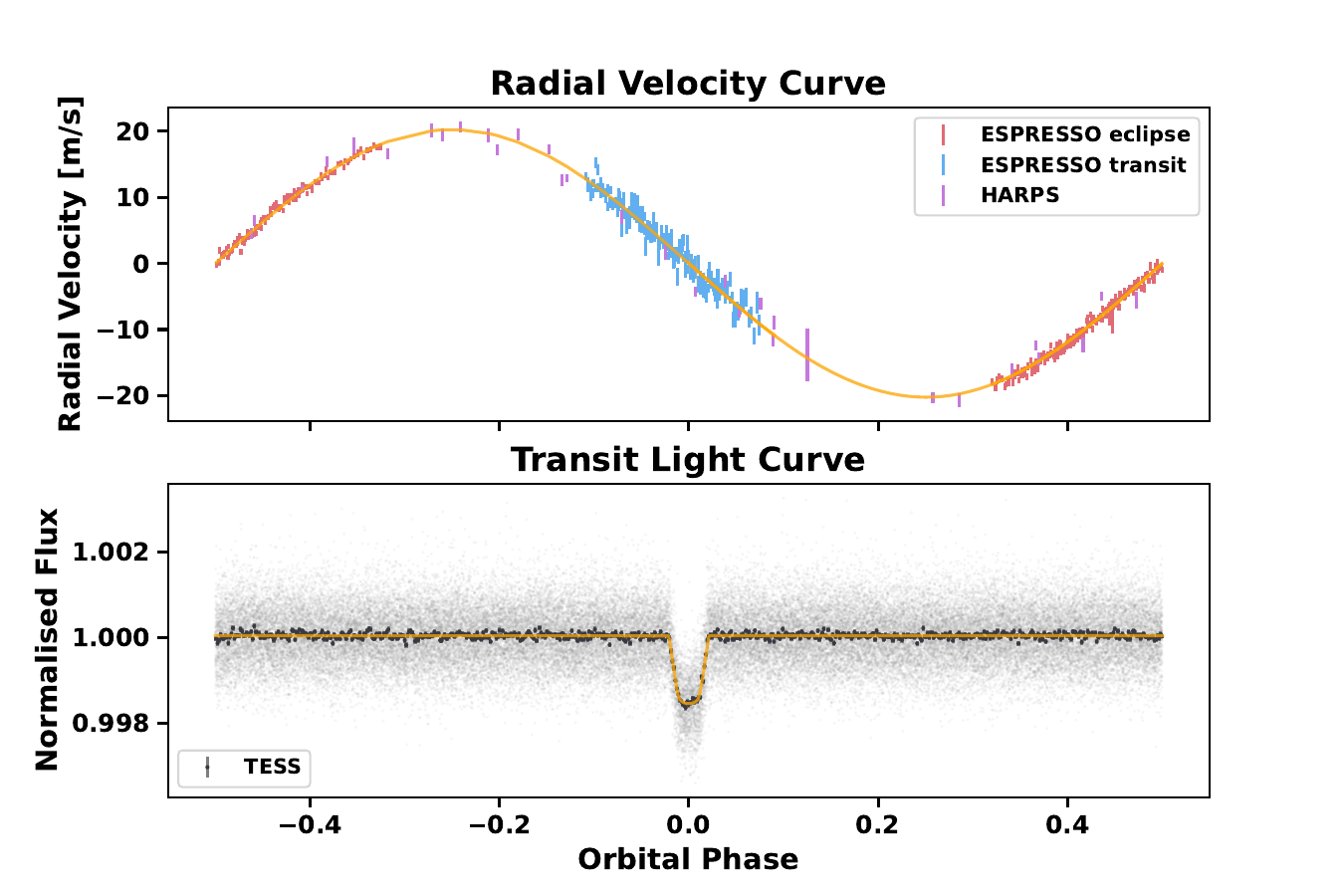}
    \caption{Top panel: phase-folded radial velocity curve comprised of radial velocity measurements from this work (ESPRESSO eclipse), ESPRESSO transit observations and HARPS. The vertical extent of each point represents the error. So that the radial velocities appear as a continuous curve, the systemic velocity and the fitted instrumental offset have been subtracted. Bottom panel: phase-folded short cadence TESS light curve (light points) with the darker points showing a binned version of the curve. In both panels, the best fit curve for a circular orbit is shown (orange).}
    \label{fig:orbitfit}
\end{figure}

\begin{figure}
    \centering
	\includegraphics[width=\columnwidth]{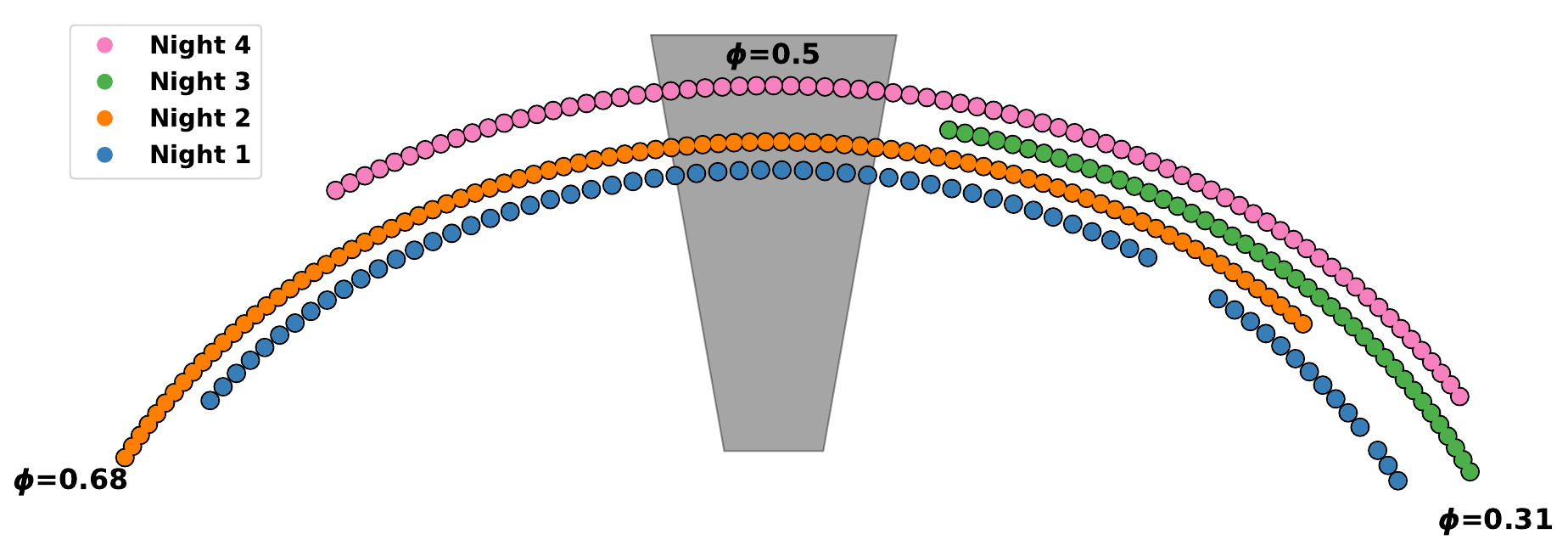}
    \caption{Phases of \lttb, assuming the ephemeris from the circular orbit fit in Table \ref{tab:orbitfit}, covered by these observations. Each circle represents an exposure and the shaded area indicates the secondary eclipse of the planet. The gaps in the observations taken on night one were due to technical difficulties and night three's observations had to be halted early due to precipitation.}
    \label{fig:orbitobservations}
\end{figure}

The eccentric orbit fit, which favours a non-zero eccentricity $e = 0.010\pm0.002$, has a slightly lower reduced $\chi^2$ indicating it is a better fit to these data. However, the Bayesian Information Criterion strongly favours the circular fit. This is because the improvement in the fit does not warrant the additional complexity that comes with adding the eccentricity as an additional parameter. Since the fit is not favoured, further radial velocity observations would be needed to robustly constrain the eccentricity of this system beyond putting a three sigma upper limit of $e < 0.016$. Figure \ref{fig:orbitfit} shows the best fit circular orbit on the phase folded radial velocities and light curve. In this work, we use the circular orbit ephemeris which is used to calculate orbital phases of \lttb\ covered by the observations shown in Figure \ref{fig:orbitobservations}. 

\begin{table*}
	\begin{center}
	\caption{Previously derived orbit parameters for \lttb\ along with those derived in this work.}
	\label{tab:orbitfit}
	\begin{tabular}{rlrll} 
		\hline
		Parameter  & Literature Ephemeris & Ref. & This work (circular) & This work (eccentric) \\
		\hline
        \hline
		Orbital Period (day) &0.792064(100$\pm$140) & 1 & 0.792064(206$\pm$117) &  0.792064(263$\pm$120)\\
		Reference transit time (BJD - 2459000) & 43.310602$\pm$0.000090 & 1 & 43.310680$\pm$0.000115 & 43.310665$\pm$0.000116\\
		Inclination, i (deg) & 76.39$\pm$0.43 & 2 & 75.03$\pm$0.05 & 75.10$\pm$0.06\\
        Eccentricity & <0.058 & 2 & 0 (fixed) & 0.010$\pm$0.002\\
		Argument of Periastron, $\omega$ (deg) & (...) & (...) & unconstrained$\dagger$ & $217^{+14}_{-17}$ \\
        Radial velocity semi-amplitude (\SI{}{\meter\per\second}) & 19.65$\pm$0.43 & 2 & 20.24$\pm$0.06 & 20.10$\pm$0.09\\
		\hline
        \hline
        Bayesian Information Criterion & (...) & (...) & 169.1 & 179.9\\
        Reduced $\chi^2$ & (...) & (...) & 1.595 & 1.580 \\
		\hline
	\end{tabular}
    \end{center}
   The Bayesian Information Criterion strongly favours the circular fit. $\dagger$ This parameter was included but remained unconstrained.\\
    \textbf{References.} (1) \citet{Edwards2023a}; (2) \citet{Jenkins2020}
\end{table*}

\section{Removing contaminating spectra}
\label{sec:datareduction}

In the time-series of observations, the position of the planetary spectral lines changes significantly due to the planet's orbital motion but the stellar lines and Earth's telluric lines remain quasi-stationary in wavelength. The HRCCS technique isolates the planet's spectrum by removing the quasi-stationary trends created by the stellar and telluric lines. However, since \lttb's spectrum is buried in the stellar spectrum at approximately the $100$ppm level \citep{Hoyer2023, Coulombe2025, Radica2025, Saha2025}, the data reduction must remove the contaminating spectra to at least this degree. In this section we discuss the standard techniques employed to achieve this level of cleaning.

\subsection{Outlier Removal}
\label{outlierremoval}

The first step of this process is to remove the outliers missed by the ESPRESSO pipeline. The outlier removal is applied to the spectral time-series of each order and each night separately in the barycentric rest frame used by the ESPRESSO pipeline. Using the barycentric rest frame for this part of the reduction minimises the need to interpolate the outliers which improves their removal. To remove the outliers, each spectrum in the time-series of a given spectral order is first continuum normalised using a fourth order polynomial fit that excludes points more than ten standard deviations from the mean. Next, the blob detection algorithm described in \citet{Kong2013} is run. It flags outliers with a threshold of $4 \sigma$ and replaces them by interpolating the surrounding non-flagged data points. The points used are both adjacent in wavelength and time, using the spectra taken before and after the outlier. Since the spectra are in the barycentric rest frame, both the stellar and telluric lines will not be stationary in wavelength. However, the Doppler shift between adjacent spectra is significantly below the instrumental resolution and thus this has minimal impact on the outlier removal. We run the algorithm either five times or until no more outliers are removed, which ever comes first. This removes the worst outliers but a visual inspection of the data reveals some have been missed. To identify these too, we transform to the stellar rest frame and remove the median spectrum of the observations taken during secondary eclipse\footnote{\label{note1}For night three all observations are used to calculate the median as there are no observations during secondary eclipse.} before restoring the data to the barycentric rest frame. This allows the blob detection algorithm to identify and replace the remaining outliers using a threshold of $2.5 \sigma$. Finally, the median spectrum is added back into the data and the continuum is restored. The reason for this last step is discussed in Section \ref{sec:modelinjection}.

\subsection{Removing Stellar and Telluric}
\label{sec:starremoval}

\begin{figure}
	\includegraphics[width=0.9\columnwidth]{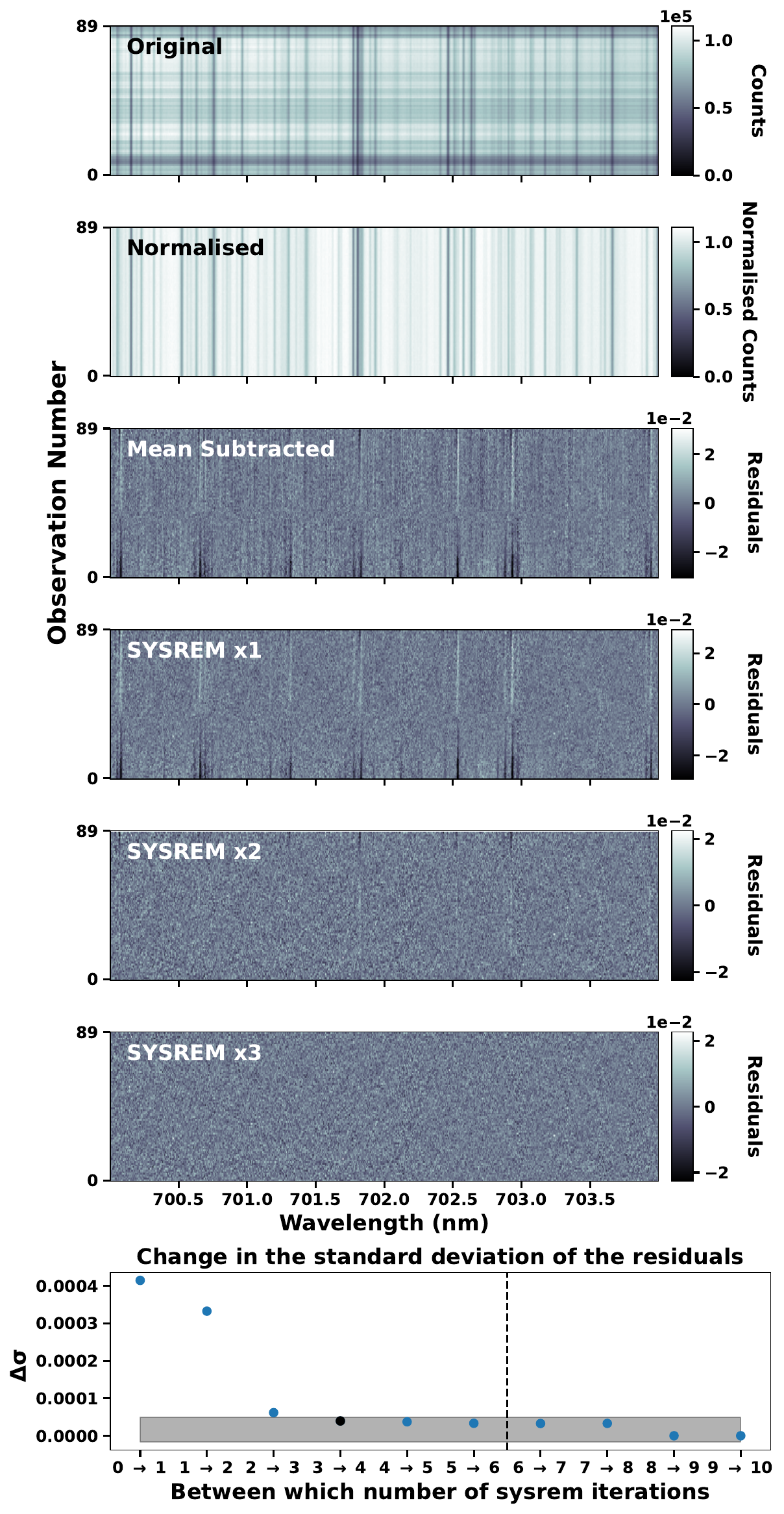}
    \caption{Schematic of the data reduction process for the spectral time-series of order 76 night two. Each of the six top panels show the time series of spectra with each row representing one spectrum in the series. Top panel: Original spectra. Second panel from the top: Spectral time-series after continuum normalisation. Third panel from the top: Spectral time-series after the mean spectrum has been subtracted. Fourth to sixth panels from the top: Spectral time-series after one, two and three \texttt{sysrem} iterations are run on the time series. Bottom panel: Difference in the standard deviation of the residuals between consecutive \texttt{sysrem} iterations as a function of the number of \texttt{sysrem} components. The shaded region highlights the one sigma deviation in the last four points (those to the right of the dashed line) and the black dot indicates the chosen number of iterations (the smaller number) as described in the text.}
    \label{fig:datareduction}
\end{figure}

To remove the contaminating stellar and telluric spectra, we perform the following on the spectral time-series of each order and each night separately since they might have different systematics. The spectra start in the barycentric rest frame but are converted into the stellar rest frame during this part of the data reduction. An example of these processes is shown in Figure \ref{fig:datareduction}.

The first step is to correct for the continuum where variations in flux are primarily due to Earth's atmosphere. This is achieved by dividing each spectrum in the time-series by a smoothed version of itself which is computed by convolving it with a box function $255$ pixels wide. This width was chosen via trial and error and visually appears to work best. Next, each spectrum in the time series is interpolated from the barycentric frame into the stellar rest frame. Once in this new frame, the median spectrum of the spectra taken during secondary eclipse\footnoteref{note1} is subtracted from each observation. 

While this removes the majority of the contamination, clear residuals from many spectral lines remain in the time-series as seen in the third panel of Figure \ref{fig:datareduction}. The dominant source of contamination in these data are the spectral lines from the star. Since the tellurics have less impact, and are primarily produced by species we are not interested in, we choose to use \texttt{sysrem} over telluric fitting models such as molecfit \citep{Smette2015, Kausch2015} to remove this contamination. The fourth to sixth panels of Figure \ref{fig:datareduction} show several iterations of this algorithm removing the remaining residuals. Since the \texttt{sysrem} algorithm focuses on agnostically removing stationary trends from the data, it is possible that it might remove some or all of the planet's spectrum. This can be the case if too many \texttt{sysrem} iterations are used, if the spectral lines change Doppler shift too slowly or are significantly Doppler smeared or rotationally broadened. These observations have been designed such that, the change in Doppler shift between observations should be large enough not to cause a significant impact and we do not expect the planet's atmospheric spectral lines to be significantly Doppler smeared\footnote{There is a small amount of Doppler smearing in night one.} or rotationally broadened ($\SI{2.7}{\kilo\meter\per\second}$ see Section \ref{sec:observations}). Some spectral lines may become intrinsically broadened when the species abundance is high or the cloud deck is low. Injection recovery tests can be used to verify if this broadening is impacting the recovery of the planet's spectrum. Therefore only the number of \texttt{sysrem} iterations might cause a problem. To select the number of \texttt{sysrem} iterations, we follow a process similar to that described in (Spring et al. in prep). The \texttt{sysrem} algorithm \citep{Tamuz2005} is run a total of ten times on the spectral time-series of each order with the result for each iteration saved. These are then used to choose the number of \texttt{sysrem} iterations. This is done by measuring the point at which the change in standard deviation of the reduced spectra in the time-series between iterations drops to within one standard deviation of the changes between the last four iterations. If there was no change in the standard deviation in the last four iterations, then the last iteration that removed some component is used instead. This is demonstrated in the last panel of Figure \ref{fig:datareduction}. 

Successful recovery of injected models, for example in Figure \ref{fig:injectionexample}, indicates that the \texttt{sysrem} algorithm is not destroying the planet's spectrum although it may still be damaging it. Ideally, one would use physics based models for the stellar spectrum like those used for the tellurics. Unfortunately, this is not currently possible and algorithms like \texttt{sysrem} are the main alternative. In Section \ref{sec:results}, we use injection recovery tests to study the constraining power of our data to different model atmospheres. Since the injected spectra undergo the same data reduction, they will be similarly damaged and so these are a realistic estimate for the sensitivity of our data given the current data reduction. 

Finally, to remove any residual slowly changing continuum, which can cause artefacts in the later analysis, each spectrum in the time-series is high pass filtered. This is achieved by subtracting a smoothed version of each spectrum computed by convolving the spectrum with a box function $255$ pixels wide. Again, this width was chosen via trial and error.

\section{Recovering the planet's spectrum}
\label{sec:recoverplanet}

While the data reduction presented in Section \ref{sec:datareduction} is able to successfully remove the stellar and telluric spectra, it is unable to remove the photon noise from the star. This noise prevents the planet's spectrum from being measured directly, even when all observations are combined. However it is still possible to extract useful information on the embedded planets spectrum by matching it to model spectra using the likelihood method. In this work, we use the likelihood metric presented in \citet{Brogi2019}.

\subsection{The likelihood method}
\label{sec:likelihood}

This method compares a Doppler shifted model spectrum, $g(\lambda_n-\lambda_s)$, to the reduced spectrum of a single observation and spectral order, $f_{obs,order}(\lambda_n)$ using the following likelihood metric:

\begin{equation}
\label{equ:likelihood}
\text{log}(L) = -\frac{N}{2} \text{log}[s_f^2 -2R(s) + s_g^2],
\end{equation}

where $s$ represents the Doppler shift of the model. In this metric, $s_f^2$ is the variance of the data and $s_g^2$ the variance of the Doppler shifted model:

\begin{equation}
s_f^2 = \frac{1}{N} \sum_{n} f_{obs,order}^2(\lambda_n); \qquad s_g^2 = \frac{1}{N} \sum_{n} g^2(\lambda_n-\lambda_s)
\end{equation}

where $g(\lambda_n-\lambda_s)$ has been interpolated to the same wavelength grid as the observations so that in both cases the sum is over the pixels, $n$, in the reduced spectrum. 

The central term in the likelihood, $R(S)$, is the cross-covariance of the data and Doppler shifted model:

\begin{equation}
R(s) = \frac{1}{N} \sum_{n} f_{obs,order}(\lambda_n)g(\lambda_n-\lambda_s).
\end{equation}

Another standard metric used in HRCCS analyses is the cross-correlation coefficient, $C(s)$, which is related to the cross-covariance via:

\begin{equation}
\label{equ:crosscorrelation}
C(s) = \frac{R(s)}{\sqrt{s^2_f s^2_g}}.
\end{equation}

Both $R(s)$ and $C(s)$ increase when the model is at the right Doppler shift and has spectral lines that match the planet's spectrum. This leads to an increase in the likelihood, $L$.

\subsection{Generation of the model spectra}
\label{sec:spectrageneration}

In the likelihood framework, we need model spectra to compare to the embedded planet's spectrum. Reflection spectra are formed of a reflected version of the stellar spectrum modulated by the planet's wavelength dependent albedo. The reflected stellar spectrum is not identical to that which we observe from the star as it experiences a different degree of rotational broadening. This is because, while we view the star rotating with its own rotation period, when viewed from the planet the star appears to be rotating with a combination of the stellar rotation period and the planet's orbital period. We use Equation 6 in \citet{Spring2022} to compute the rotational broadening of the reflected stellar spectrum for \lttb. This yields a broadening of $\nu_{ref} \sim \SI{60}{\kilo\meter\per\second}$. Therefore \lttb's spectrum will not contain any deep stellar spectral lines and we model the stellar spectrum as a continuum. To model the modulation of the reflected stellar spectrum caused by \ltt's atmosphere, we use \texttt{PICASO} \citep{Batalha2019a} to compute the planet-star contrast ratio. The atmospheric spectral lines are broadened only by the planetary spin. Assuming tidal locking, this results in a broadening of approximately $\SI{2.7}{\kilo\meter\per\second}$ which is below the instrument resolution of $\SI{4.3}{\kilo\meter\per\second}$ and so it has been neglected when computing the planet-star contrast ratio. The model of the planet's spectrum is then created by multiplying the modelled stellar spectrum with the generated planet-star contrast ratio. 

Our ESPRESSO spectra have $R\sim70,000$ but \texttt{PICASO} needs opacities several times this resolution to accurately reproduce the spectrum. We elect to use opacities with $R\sim500,000$ which are listed in Table \ref{tab:opas}. Since the data reduction uses a high pass filter, the same filtering must be applied to the models so they can match the real embedded spectrum. We achieve this high pass filtering by subtracting a $R=2,000$ version of the model spectrum from the full $R=70,000$ version. This lower resolution was chosen so that the model spectrum would be smoothed by a filter of similar width to the data (see Section \ref{sec:datareduction}).

To generate the planet-star contrast ratio, \texttt{PICASO} requires the atmospheric temperature-profile to be specified along with the volume mixing ratios (VMR) of the species in the atmosphere and a model of the clouds. The parameters of all the models used in this work are summarised in Table \ref{tab:models} and discussed in details in the following sections.

\subsubsection{Spectral models based on JWST NIRISS/SOSS observations}
\label{sec:ltttemplates}

We use the information provided by previous JWST NIRISS/SOSS observations to construct high spectral resolution model spectra. Examples of the temperature-profile, species VMRs and cloud profile used by \texttt{PICASO} to compute these `self-consistent' model spectra are shown in Figure \ref{fig:spectra_inputs} and summarised in Table \ref{tab:models}. 

We compute three sets of `self-consistent' model spectra, each for a different segment of the planet's dayside. These are eastern-dayside, central-dayside (hereafter dayside) and western-dayside as defined in \citet{Coulombe2025}. Each segment has a different temperature-profile for which we use the mean measured from JWST NIRISS/SOSS phase curve observations \citep[][shown here in Figure \ref{fig:spectra_inputs}; T.1 in Table \ref{tab:models}]{Coulombe2025_dataset}. There is a large uncertainty in the temperature-profile at low pressures (less than $10^{-6}$ bar) as the the JWST NIRISS/SOSS phase curve data are not particularly sensitive to this region of the atmosphere. Therefore, the low pressure part of the temperature-profile has tended towards the centre of the priors used by \citet{Coulombe2025}, which results in a hot upper atmosphere. To explore how much our results might change if the upper atmosphere were cooler, as is suggested by the lack of evidence for a thermal inversion \citep{Dragomir2020}, we also create a similar set of models where the temperature-profile has been made isothermal above $10^{-5}$ bar (T.2 in Table \ref{tab:models} and shown in \ref{fig:spectra_inputs}). This results in a total of six different temperature-profiles, two for each of the three atmospheric segments.

For all six temperature-profiles we assume equilibrium chemistry and compute the VMRs using \texttt{easychem} \citep{Lei2024} for a range of metallicities from $0.1\times$ solar to $1000\times$ solar (Ch.1 in Table \ref{tab:models}). The VMRs computed for the $10\times$ solar metallicity model are shown in Figure \ref{fig:spectra_inputs}. As discussed in Section \ref{sec:introduction}, it is currently uncertain if TiO and VO are present at their equilibrium abundances or if they are depleted. We therefore create variants of these models with the VMR of TiO set to zero (Ch.2 in Table \ref{tab:models}) and with the VMRs of TiO and VO set to zero (Ch.3 in Table \ref{tab:models}) for the same range of metallicities and for all six temperature-profiles. The VMR profiles of some species do differ significantly between the original (T.1) and modified (T.2) temperature-profiles as a result of changes in the reaction rates modelled within the \texttt{easychem} chemical network, including notably the profiles of TiO and VO. 

Previous works have favoured the presence of Mg$_2$SiO$_4$ and MgSiO$_3$ clouds over those of other types \citep{Hoyer2023, Radica2024a, Coulombe2025, Radica2025}. To model these clouds, we use the \texttt{virga} module \citep{Batalha2025} which implements the \citet{Ackerman2001} cloud model to compute the extinction and scattering parameters for a set of condensates based on the vertical mixing, $k_{zz}$, and sedimentation efficiency, $f_{sed}$. This method is similar to that used in \citet{Saha2025}. We use \texttt{virga} to specify appropriate parameters for only Mg$_2$SiO$_4$ and MgSiO$_3$ clouds \citep{Scott1996} for each segment assuming $k_{zz} = 10^9$, $f_{sed} = 1$ and that the particle sizes follow a log normal distribution with $\sigma = 2$. These parameters are consistent with \citet{Radica2025}. The optical depth profile of the combined Mg$_2$SiO$_4$ and MgSiO$_3$ clouds are shown in Figure \ref{fig:spectra_inputs} where the top of the clouds are at an altitude of approximately $10^{-4}$ bar for the western-dayside and dayside models and $10^{-1.5}$ bar for the eastern-dayside model. We note that, the vertical extent and height of these clouds can be changed by varying $k_{zz}$, $f_{sed}$ or the temperature-profile. The altitude of the clouds are not strongly constrained with the JWST NIRISS/SOSS data as the uncertainty in the intersect point between the temperature-profile and the cloud species condensation curve that spans a large fraction of the atmosphere (see Figure \ref{fig:spectra_inputs}). Given the complex way the resulting cloud model is influenced by the input parameters, we elect to use the simpler models described in Section \ref{sec:searchgridtemplates} for exploring different cloud scenarios and keep the cloud model fixed for the `self-consistent' models. Additionally, since the modified temperature-profile does not significantly alter the cloud profile, we opt to use the same cloud model for both the original (T.1) and modified (T.2) temperature-profiles.

\begin{figure*}
    \centering
	\includegraphics[width=0.7\textwidth]{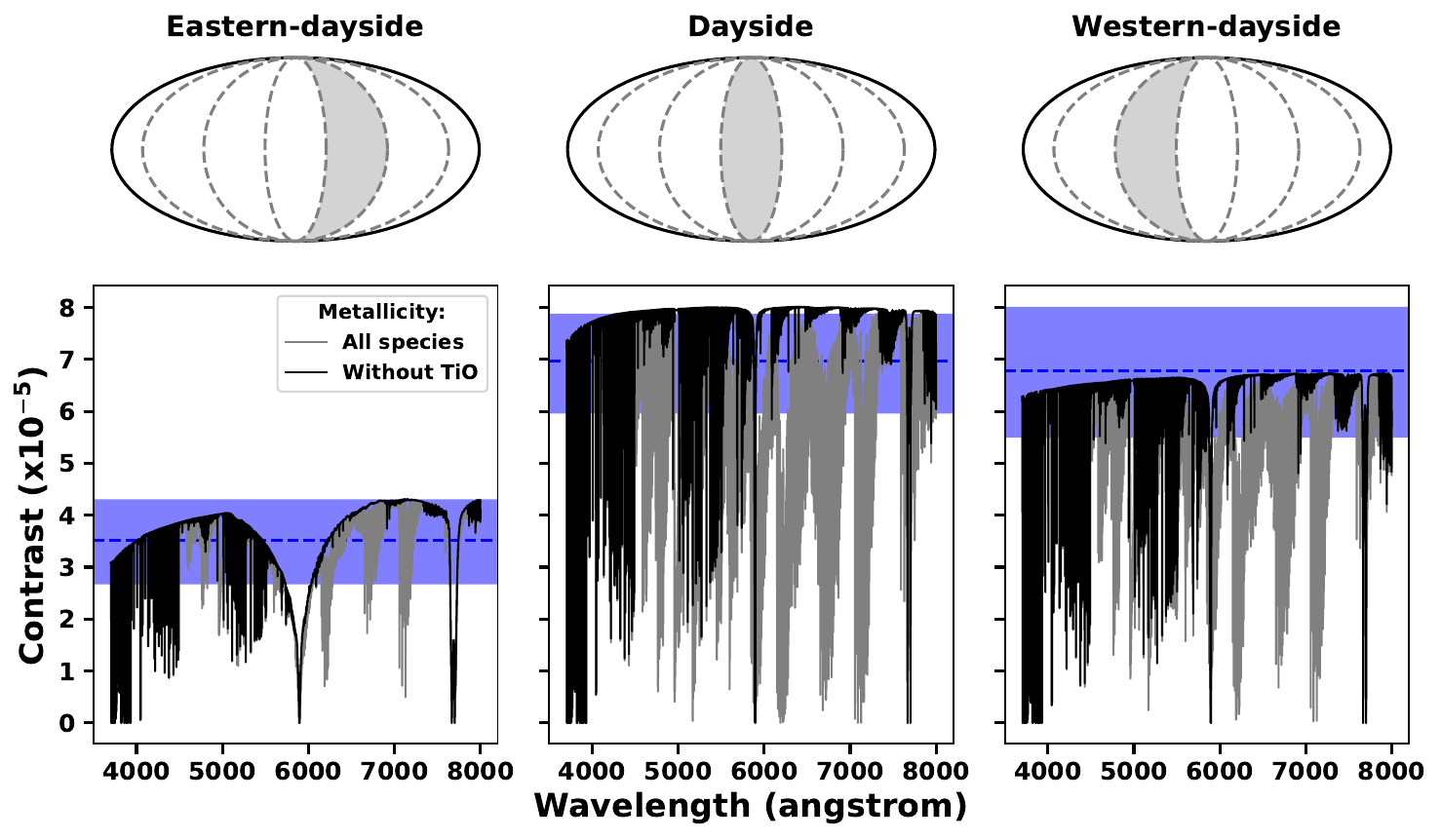}
    \caption{High spectral resolution ($R=70,000$) models computed using the eastern-dayside, dayside, and western-dayside mean temperature-profiles from \citet{Coulombe2025} (T.1) assuming $10\times$ solar metallicity. Each panel shows two spectra, one that assumes equilibrium chemistry for all species (Ch.1) and one that removes TiO from the atmosphere (Ch.2). The contrast ratio measured by \citet{Coulombe2025} (see Fig. 3 of their work) for each segment is shown as the dashed line with the shaded region indicating the one sigma errors.}
    \label{fig:ltttemplates}
\end{figure*}

Figure \ref{fig:ltttemplates} shows an example `self-consistent' model spectrum for each segment of the planet's dayside for the original temperature-profile (T.1) and $10\times$ solar metallicity. Each panel shows two versions of the spectra, one with all chemical species (Ch.1) and one with TiO removed (Ch.2). From this, it is clear that TiO is responsible for a large number of the spectral lines. This Figure includes the contrast ratio and its error from Fig. 3 of \citet{Coulombe2025} which we reproduce using \citet{Coulombe2025_dataset}. This demonstrates that these spectra are largely consistent with these previously measured values.

\subsubsection{Free-chemistry single species models}
\label{sec:searchgridtemplates}

The `self-consistent' models assume equilibrium chemistry, however, there is uncertainty in which species are actually present at high altitudes. \citet{Dragomir2020} found a non-inverted temperature-profile which might suggest a lack of TiO and VO at high altitudes and \citet{Zhou2025} suggested that the modelling difficulty they encountered may be the result of some form of disequilibrium chemistry. Disequilibrium chemistry has been seen in giant planet atmospheres before \citep[e.g.][]{Fortney2020, Sing2024a} so it is possible that this atmosphere might not be in equilibrium. Thus we create sets of `single-species' model spectra to agnostically search for the presence of individual species, while recognising that the detection strength of such models will be lower that for multi-species models due to the loss of signal caused by excluding the spectral lines of other species. Table \ref{tab:models} summarises the inputs for these the models. 

In reflected light, the temperature-profile influences the spectrum primarily by setting the chemical equilibrium and condensation points for clouds. However, if both the VMRs and cloud properties are set independently of the temperature-profile then the spectrum is not strongly influenced by its parametrisation as the opacities are not strongly temperature dependent. Therefore, for simplicity, we assume an isothermal temperature-profile set to $1800$~K and set the VMRs and cloud properties independently.

For the chemical composition, we assume a H/He dominated atmosphere with a trace amount of a single additional species. We choose TiO, VO, Fe, MgH, FeH, H$_2$O and AlH as these species have a large number of spectral lines in the ESPRESSO wavelength range and produce spectral features at feasible abundances. Therefore these represent the best chance for individual detection. The VMR of the trace species is assumed to be constant within the atmosphere and Table \ref{tab:models} summaries the values assumed for each species.

Next, we model a single grey cloud deck with a free choice of the altitude, vertical extent, optical depth, asymmetry parameter and single scatter albedo. We assume a fixed vertical extent of one order of magnitude in pressure and an optical depth equal to ten. The single scatter albedo, $\omega_0$ and asymmetry parameter $g_0$ are set such that the continuum matches the pre- and post-eclipse contrast measure by JWST NIRISS/SOSS \citep{Coulombe2025_dataset}. That is $\omega_0=0.885$, $g_0=0.0$ for pre-eclipse (C.1 in Table \ref{tab:models}) and $\omega_0=0.995$, $g_0=0.0$ for post-eclipse (C.2 in Table \ref{tab:models}). This creates two sets of `single-species' models for each species, one for each side of the eclipse. The altitude of the top of the cloud deck is varied in steps of an half order of magnitude in pressure ranging from $10^{0.5}$ to $10^{-5}$ bar. 

For each species we create a grid of models that vary the VMR of that species and the height of the cloud deck. The spectral lines shapes of the trace species vary and can be used to identify the presence of a given species in our data. 

\begin{table*}[]
\caption{Summary of the inputs to the model spectra.}
\label{tab:models}
\addtolength{\tabcolsep}{-0.2em}
\begin{tabular}{lll}
\hline
Component & Self-consistent & Free-chemistry \\
\hline
\hline
\begin{tabular}[c]{@{}l@{}}Temperature\\ Profile\end{tabular} & \begin{tabular}[c]{@{}l@{}}T.1) mean profile measured by JWST NIRISS/SOSS\\ for orbital phases in Table \ref{tab:phasesplit} (see also Figure \ref{fig:spectra_inputs})\\ T.2) As T.1) but isothermal above $10^{-5}$ bar \\ \end{tabular} & Isothermal: T=1800K  \\
\hline                                     
Chemistry & \begin{tabular}[c]{@{}l@{}}Ch.1) Equilibrium chemistry for metallicities \\ \quad $0.1\times$, $1\times$, $10\times$, $100\times$, $1000\times$ solar\\ Ch.2) Same as Ch.1) but with TiO removed \\ Ch.3) Same as Ch.1) but with TiO and VO removed\end{tabular} & \begin{tabular}[c]{@{}l@{}}H/He atmosphere with a uniform volume \\ mixing ratio of a single species, log10(VMR$_{x}$):\\ $x\rightarrow$ TiO: \{-5,-6,-7,...,-15\} \\ $x\rightarrow$ VO: \{-5,-6,-7,...,-15\} \\ $x\rightarrow$ Fe: \{-2,-3,-4,...,-7\} \\ $x\rightarrow$ MgH: \{-4,-5,-6,...,-14\} \\ $x\rightarrow$ FeH: \{-5,-6,-7,...,-13\} \\ $x\rightarrow$ H$_2$O: \{-1,-2,-3,...,-8\} \\ $x\rightarrow$ AlH: \{-5,-6,-7,...,-13\} \end{tabular}  \\
\hline
Clouds & \begin{tabular}[c]{@{}l@{}}Mg$_2$SiO$_4$ and MgSiO$_3$ clouds modelled with \texttt{virga}\\ assuming $k_{zz} = 10^9$, $f_{sed} = 1$ and the above temperature-profiles\end{tabular} & \begin{tabular}[c]{@{}l@{}}Grey cloud deck with an optical depth of 10\\ spanning one order of magnitude in pressure. \\ \\ Top of the cloud deck pressure, P$_{\text{top}}$:\\ $10^{\{0.5,0,-0.5,...,-5\}}$ bar \\ \\ Cloud scattering properties: \\ \begin{tabular}[c]{ccc} \hline Parameter & Pre-eclipse (C.1) & Post-eclipse (C.2) \\ \hline\hline $\omega_0$ & $0.885$ & $0.995$ \\ $g_0$ & $0.0$ & $0.0$ \\  \end{tabular} \end{tabular} \\
\hline
\end{tabular}
\end{table*}

\subsection{Combining all the data}
\label{sec:combiningdata}

The likelihood shown in Equation \ref{equ:likelihood} of Section \ref{sec:likelihood} and the cross-correlation coefficient in Equation \ref{equ:crosscorrelation} are computed for a range of model Doppler shifts from $\SI{-500}{\kilo\meter\per\second}$ to $\SI{500}{\kilo\meter\per\second}$ in steps of $\SI{0.8}{\kilo\meter\per\second}$ on the reduced data of each observation and spectral order separately. This step size is chosen as it is approximately equivalent to the velocity spacing of each pixel in the spectrum. This is smaller than the resolution element, $\SI{4.3}{\kilo\meter\per\second}$, as ESPRESSO over-samples the spectrum such that instrumental resolution is equivalent to 5.5 pixels on the detector. The planet's spectrum is too faint to show a statistically significant increase in the likelihood or cross-correlation coefficients for each observation and order in isolation. To identify the planet's spectrum, all the orders and observations must be combined. 

The combination process is the same for both the likelihood and cross-correlation values, however, for clarity only the likelihood values are mentioned in the following explanation. First, the likelihoods of each order are summed resulting in one likelihood value for each observation and model Doppler shift. Combining the observations is more complicated as the planet's velocity is changing and the planet's spectrum itself is changing as evidenced by differences seen in the eastern-dayside, dayside and western-dayside models shown in Figure \ref{fig:ltttemplates}. To account for the changing spectrum, the observations can be split up based on the orbital phases they cover. To account for the orbital motion during the selected observations, the likelihood values are summed along a variety of orbital paths defined by:

\begin{equation}
V_{planet}(t) = V_{sys} + K_p \sin( 2 \pi \phi(t) )
\end{equation}

where $\phi$ is the a priori known phase of the orbit and the systemic velocity, $V_{sys}$, and orbital velocity, $K_p$, both range from $\SI{-500}{\kilo\meter\per\second}$ to $\SI{500}{\kilo\meter\per\second}$ in steps of $\SI{0.8}{\kilo\meter\per\second}$. 

This process creates a $V_{sys}$-$K_p$ map, a 2D colour map indicating how the sum of either the likelihoods or cross-correlation coefficients varies when the data is combined assuming different orbits for the planet. Since the presence of the planet's spectrum creates an increase in the likelihood and cross-correlation coefficient, the sum will be higher when all these small increases are summed together using the correct orbit. This leads to a peak in the $V_{sys}$-$K_p$ map at the location where the systemic and orbital velocities match that of the planet. This will not be the only peak in the map as noise can cause fluctuations in the sum creating a series of peaks and troughs over the whole map. These might also show correlated structure if residuals from the Earth's tellurics or the stellar spectrum remain in the data. Alternatively this structure might come from the planet itself if the signal is strong and the autocorrelation function of the planet's spectrum has alias peaks. Therefore these maps must be converted via some metric to determine if the peak corresponding to the planet's signal is significantly above the noise level in the map.

\subsection{Measuring the significance of a detection}
\label{sec:detectionsignificance}

For maps of cross-correlation coefficients, the detection strength of a signal is typically measured by dividing the map by its noise, converting the coefficients into signal-to-noise. The noise is taken to be the standard deviation of the map excluding the region around the expected location of the planet's signal and a signal-to-noise of five or more if typically needed for a robust detection. The left panel in Figure \ref{fig:nondetection} shows an example $V_{sys}$-$K_p$ map in terms of signal-to-noise which has been calculated from the cross-correlation coefficients by dividing each row by its standard deviation excluding a $\SI{60}{\kilo\meter\per\second}$ swath around $V_{sys}=\SI{0}{\kilo\meter\per\second}$ to exclude any possible planetary signal. There is no signal from the planet in this map which would appear as a peak at $V_{sys}=\SI{0}{\kilo\meter\per\second}$ and $K_p= \SI{224}{\kilo\meter\per\second}$. 

For map of likelihoods, the detection strength is determined by converting the likelihood values to a significance. In this work, we use the process described in more detail in Appendix D of \citet{Pino2020}, which has often been used in HRCCS exoplanet characterisation \citep{Lafarga2023, vanSluijs2023, Dash2024, Parker2025}. This method uses Wilks' theorem \citet{Wilks1938} which states that the test statistic (Equation \ref{equ:wilks}) follows a $\chi^2$ distribution in the limit of large sample sizes. This distribution has degrees of freedom equal to the difference in degrees of freedom between the null hypothesis, that the planet's orbit is described by a specific pair of $(V_{sys, \text{null}}, K_{p,\text{null}})$, and the alternative hypothesis, where the orbital parameters are allowed to vary. In this case this means the $\chi^2$ distribution has two degrees of freedom.   

The survival function of the $\chi^2$ provides the probability that the likelihood measured for an alternative set of parameters $(V_{sys, \text{alt}}, K_{p,\text{alt}})$ would occur by chance if the parameters for the null model $(V_{sys, \text{null}}, K_{p,\text{null}})$ were the true values. If the alternative models all have small probabilities, this implies that the null model represents an increase in the likelihood above the level expected for noise. To quantify this in an intuitive way, we convert the probabilities to significances using a Gaussian distribution. 

We compute the test statistic where the null model, $(V_{sys, \text{null}}, K_{p,\text{null}})$, has the maximum likelihood, $L_{\text{max}}$, in the map:

\begin{equation}
\label{equ:wilks}
\lambda = 2 [\text{log}(L_{\text{max}}) - \text{log}(L)]  .  
\end{equation}

The null model may not necessarily have the expected $V_{sys}=\SI{0}{\kilo\meter\per\second}$ and $K_p= \SI{224}{\kilo\meter\per\second}$ for \lttb. If it similar, it could simply be due to fluctuations in the noise. However, if the null parameters are significantly different, then this means the planet has not been detected as its signal is below the noise level. 

Converting this test statistic to significance as described creates a $V_{sys}$-$K_p$ map where the highest peak has a significance of zero and all other points have positive significance. To determine if this highest peak is significantly above the noise level in the map we follow \citet{Parker2025}. So that this map has a similar behaviour to that of the more traditional cross-correlation maps where the planet's signal would form a positive peak, we convert the significances in the map using the following metric:

\begin{equation}
\Delta\sigma_{V_{sys}, K_p} = \sigma_{\text{average}} - \sigma_{V_{sys}, K_p}.  
\end{equation}

In this work we consider $\Delta\sigma > 5$ a robust detection. The right panel in Figure \ref{fig:nondetection} shows the same map as the left panel but now in terms of likelihoods converted to significances. This map also shows no signal from the planet. In what follows we choose to only use the likelihood maps and their significances as this measure is more robust than the cross-correlation signal-to-noise. This is because the latter can be dependent on the region selected for calculating the standard deviation. The likelihood version is also easier to compare and combine with lower resolution methods.

\subsection{Defining an equivalent uniform volume mixing ratio}
\label{sec:EUvmr}

The signal-to-noise or significance at which a spectrum is recovered using high resolution cross-correlation spectroscopy is strongly dependent on the number of spectral lines and their depth. The exact number and depth of lines is a complex function of the atmospheric parameters however, it is useful to understand when these numbers might be similar for different models. To achieve this, we define the equivalent uniform volume mixing ratio, EU$_{\text{VMR}}$, as the VMR that, for a uniform profile, would result in the same column mass of a given species above the clouds as a more complex profile. 

This parameter is a good metric for determining when two models have a similar number and depth of spectral lines from a given species because the depth of the spectral lines is related to their optical depth, $\tau$, which is give by: 

\begin{equation}
    \tau = \int X \frac{\kappa}{g} dP
\end{equation}

where $X$ is the mass fraction of the species, $\kappa$ its opacity, $g$ the gravitational acceleration and $P$ the atmospheric pressure. Assuming the opacity is constant with pressure and that $g$ is also approximately constant, which is valid for the upper parts of the atmosphere. Then the optical depth and therefore depth of the spectral lines is proportional to the integrated mass fraction which is proportional to the column mass of the species. Therefore, the line depths approximately proportional to the column mass of the species. This integral needs to be limited to the atmosphere above the clouds if the clouds are optically thick as the atmosphere beneath them does not contribute to the spectrum.

\subsection{Model injection to test data sensitivity}
\label{sec:modelinjection}

The sensitivity of these data to different types of planetary spectra can be determined by testing whether or not an injected model is recovered using this analysis. To be realistic, the injected spectrum must be processed in the same way as the real spectrum. Therefore it must be injected into the data as early as possible in the analysis. As discussed in Section \ref{outlierremoval}, the first few steps in the data reduction involve continuum normalisation, subtracting the median spectrum and removing outliers. The first two of these may affect the embedded planet's spectrum but the outlier removal itself is unlikely to have a strong impact. Unfortunately, the blob detection algorithm used for the outlier removal is very computationally expensive. Thus we inject the model after this step. However, to account for the effect continuum normalisation and subtracting the median spectrum might have on the injected spectrum, the median spectrum is added back into the data and continuum restored before the model is injected. Therefore, when the data reduction proceeds, the injected spectrum is treated in largely the same way as any real spectrum in the data.

The injection is achieved by first measuring the continuum level of each order of each observation. This is done by fitting a linear trend to the spectrum excluding any deep spectral lines which are defined as those points lying more than three sigma away from the linear trend. This trend is assumed to represent the flux of the star in these data and thus the planet's spectrum would be equivalent to this trend multiplied by a model of the planet-to-star contrast ratio. This model of the contrast ratio spectrum must be first Doppler shifted to the planet's velocity but with the orbital velocity taking the opposite sign to ensure it is not injected on top of any real planetary spectra in the data. It is then multiplied by the linear trend, creating a model of the planet's flux in terms of data units, and added into the data. The data reduction and analysis then proceed as before.

\section{Results}
\label{sec:results}

\subsection{A non-detection of the reflected light of \lttb}
\label{sec:nondetection}

\begin{figure}
	\includegraphics[width=\columnwidth]{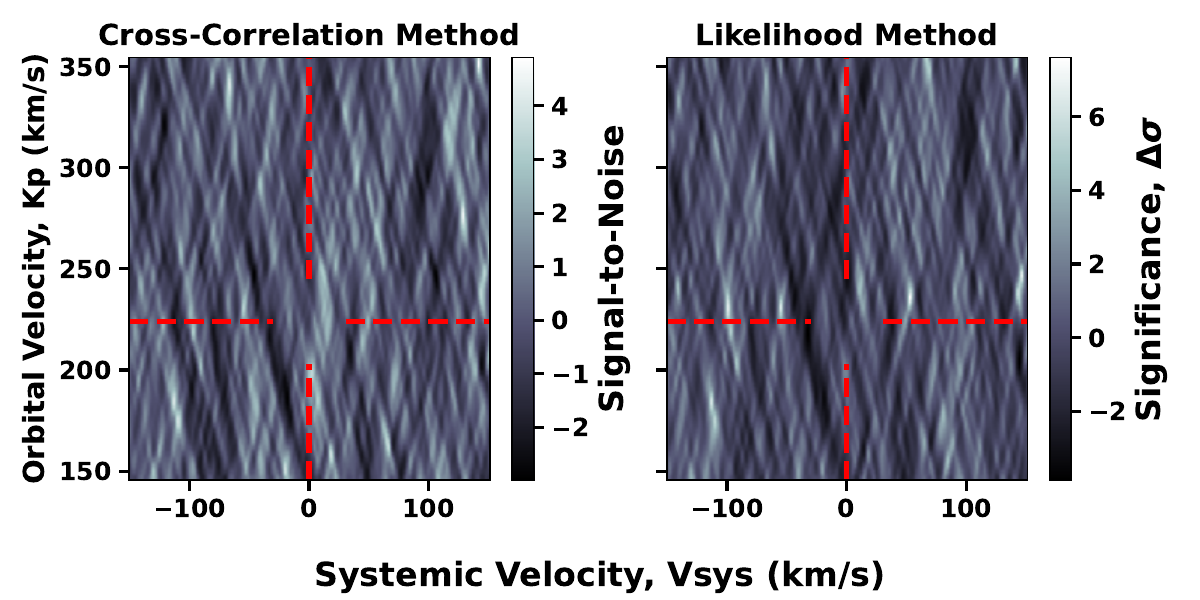}
    \caption{$V_{sys}$-$K_p$ maps for the `self-consistent' $100\times$ solar dayside model for the original temperature-profile (T.1). Left panel: $V_{sys}$-$K_p$ map in terms of signal-to-noise computed from the cross-correlation coefficients. Right panel: $V_{sys}$-$K_p$ map in terms of the significance computed from the likelihood (see Section \ref{sec:detectionsignificance}). Neither map shows a significant peak at the known orbital parameters of the planet ($V_{sys}=\SI{0}{\kilo\meter\per\second}$, $K_p= \SI{224}{\kilo\meter\per\second}$) indicated by the crosshairs.}
    \label{fig:nondetection}
\end{figure}

As discussed in Section \ref{sec:ltttemplates}, the `self-consistent' spectral models were generated for three different segments of the planet's dayside: eastern-dayside, dayside and western-dayside. These are observable at different points in the planets orbit, therefore we split our reduced data up into three sets based on the orbital phase of \lttb\ (see Table \ref{tab:phasesplit}). On each set of observations, we perform the likelihood analysis using only the models for the corresponding dayside segment. We observe no robust detection, with $\Delta\sigma>5$ or signal-to-noise > 5, of \lttb's reflected light spectrum for any of the `self-consistent' spectral models. Figure \ref{fig:nondetection} shows the $V_{sys}$-$K_p$ maps for $100\times$ solar dayside model for the original temperature-profile both in terms of cross-correlation signal-to-noise and likelihood significance, $\Delta\sigma$.

Similar to the `self-consistent' spectral models, the `single-species' models discussed in Section \ref{sec:searchgridtemplates} use two different cloud models (C.1 and C.2 in Table \ref{tab:models}) that match the albedo measured by JWST NIRISS/SOSS for \lttb's eastern (pre-eclipse) and western (post-eclipse) hemispheres. We therefore split the observations into pre- and post-eclipse and run the analysis only for the models corresponding to that part of the orbit. While these models contain only one species, and so would not perfectly match \lttb's real spectrum, a detection with these models can be used as a start point for an atmospheric retrieval which would reveal the planet's spectrum in more detail. Unfortunately, we observe no robust detection ($\Delta\sigma>5$ or signal-to-noise > 5) of \lttb's reflected light spectrum for any of the `single-species' spectral models. 

A lack of a robust detection does not imply that the planet is not reflective, simply that its high resolution reflection spectrum is not detectable with the current data as the HRCCS technique relies on the presence of a large number of deep spectral lines.

\subsection{A word of caution on false-positives}
\label{sec:falsepositive}

\begin{figure}
	\includegraphics[width=\columnwidth]{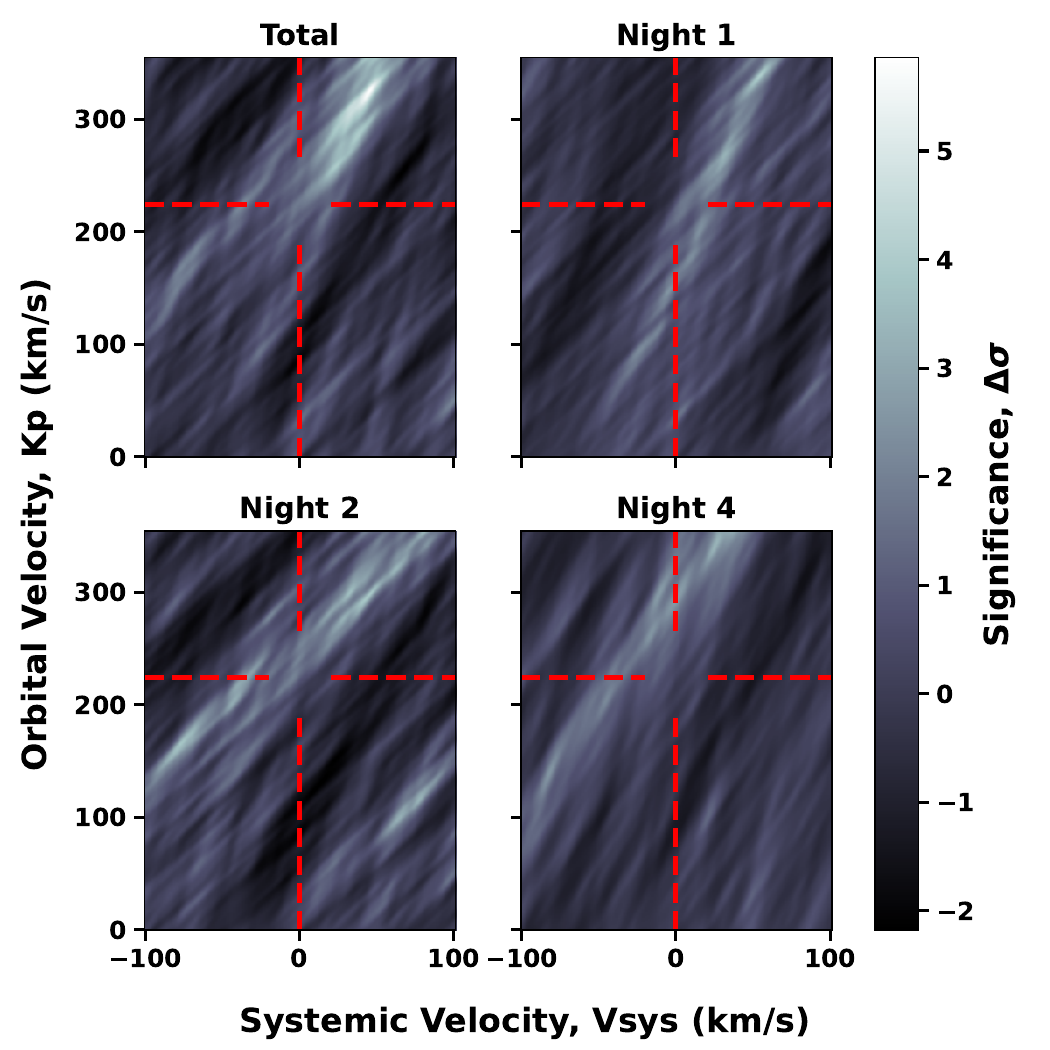}
    \caption{Top left panel: $V_{sys}$-$K_p$ map for for the `single-species' Fe model with a VMR of $0.01$ and a cloud top pressure of $10^{-1.5}$ bar for all post-eclipse data containing a significant false positive signal. The red crosshairs indicate the expected location of the planet's signal. The remaining three panels show the corresponding maps for each of the observing nights individually. There is no map for night three as no post-eclipse data was taken.}
    \label{fig:falsepositive}
\end{figure}

The $V_{sys}$-$K_p$ map for the post-eclipse data with the `single-species' Fe model that has a VMR of Fe equal to $10^{-2}$ and a cloud top pressure of $10^{-1.5}$ bar is shown in Figure \ref{fig:falsepositive}. In this map, there is a significant peak at $V_{sys} = \SI{38}{\kilo\meter\per\second}$ and $K_p = \SI{286}{\kilo\meter\per\second}$ when all the observations are combined. This is offset from the expected orbit, which is represented by $V_{sys}=\SI{0}{\kilo\meter\per\second}$ and $K_p= \SI{224}{\kilo\meter\per\second}$, by an amount that is not easily explained by one or a combination of atmospheric dynamics \citep[see Figure 9 in ][ and references therein]{Snellen2025}, errors in the ephemeris \citep[see Section \ref{sec:orbit} and][]{Meziani2025} or limb asymmetry \citet{Hoeijmakers2024}. 

Normally, we use a detection threshold of $\Delta\sigma=5$ for determining when a signal is significantly detected and while the signal here has $\Delta\sigma \approx 6.7$, it is a false-positive. This is because, due to the modelled atmospheric conditions in the lower atmosphere, the iron lines present in this and similar model spectra have broad profiles. In most HRCCS analyses, the width of the spectral lines is approximately the same as the resolution element of the observations. When the spectral lines are significantly broader than the resolution element, this Doppler shift sampling creates correlations in the noise present in the $V_{sys}$-$K_p$ map. This lowers the standard deviation of the noise, which results in more significant false-positive peaks. A visual inspection reveals that most of the spectral lines have been broadened meaning we cannot simply exclude the broad lines from the template. \citet{Parker2024} discussed how downsampling the velocity resolution in the analysis to the width of the broadened model template can mitigate correlated noise. For this model, the full-width half-maximum of the autocorrelation function is approximately $\SI{26}{\kilo\meter\per\second}$. We therefore downsample the velocity resolution by using only every 32nd Doppler shift from the original analysis. When we create the $V_{sys}$-$K_p$ map with this reduced sampling, the peak shown in Figure \ref{fig:falsepositive} reduces in significance to $\Delta\sigma<5$.

\subsection{Constraining the atmospheric parameters of \lttb\ via injection tests}
\label{sec:recoverytest}

To learn more from the non-detection, we can perform two types of injection tests to constrain the properties of \lttb. First, we could assume the planet's spectrum contains specific set of spectral lines of known depth and vary the broadband albedo of the model such as in \citet{Spring2022}. This would constrain the broadband albedo however, it is difficult to implement for \lttb\ since we do not have enough information on the deep spectral lines present. This is because the stellar lines cannot be used in this analysis due to strong rotational broadening and prior constraints do not allow us to confidently predict which atmospheric spectral lines will be present. Therefore, we elect not to perform this test, as the results will be highly dependent on which model we choose to use for \lttb's spectrum and the albedo has already been measured at low spectral resolution. 

\begin{figure}
    \centering
	\includegraphics[width=0.9\columnwidth]{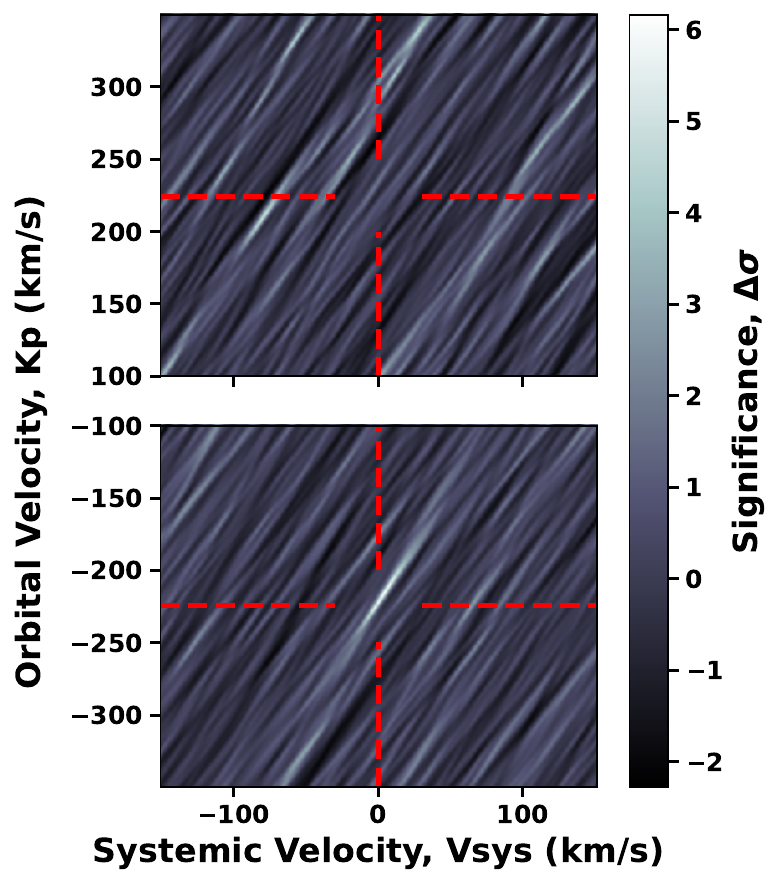}
    \caption{Example of an injection-recovery test. Top panel: $V_{sys}$-$K_p$ map when the true observations are compared to the `self-consistent' $100\times$ solar western-dayside model with the original temperature-profile. Bottom panel: $V_{sys}$-$K_p$ map when the same model is injected into the data at a negative orbital velocity (note the reversed y-axis). The expected location of the signal is indicated by the crosshairs in both panels.}
    \label{fig:injectionexample}
\end{figure}

The second type of injection test assumes instead that the broadband albedo is known and that the type and depth of the spectral lines are varied. This type of test can put constraints on the abundances of the species in \lttb's atmosphere and on altitude of the cloud deck. We elect to perform this type of test. Models are injected into the data prior to the main data reduction as described in Section \ref{sec:modelinjection} with the analysis proceeding as described in Sections \ref{sec:datareduction} and \ref{sec:recoverplanet}. Figure \ref{fig:injectionexample} shows an example of the $V_{sys}$-$K_p$ map from an injection-recovery test in which the injected model is recovered at high significance. This means that this model cannot represent the planet's spectrum as no detection was made with the original data. 

\subsubsection{Sensitivity to the `self-consistent' spectral models}
\label{sec:lttrecovery}

Section \ref{sec:ltttemplates} introduces the `self-consistent' spectral models which use atmospheric parameters measured by JWST NIRISS/SOSS for three segments of the planet's dayside: eastern-dayside, dayside and western-dayside. These different segments are observable at the orbital phases given in Table \ref{tab:phasesplit}.

For each type of temperature-profile (T.1 and T.2 in Table \ref{tab:models}) and chemistry (Ch.1, Ch.2 and Ch.3 in Table \ref{tab:models}) we inject the three different segment models into the observations of the corresponding orbital phases. The data is reduced as described in Section \ref{sec:datareduction} and the reduced observations split up into three sets based on which of the three segment models were injected. The likelihood analysis is performed on each of the three sets as described in Section \ref{sec:recoverplanet} using the injected spectrum as the model in the likelihood analysis ($g$ in equation \ref{equ:likelihood}).

\begin{figure}
	\includegraphics[width=\columnwidth]{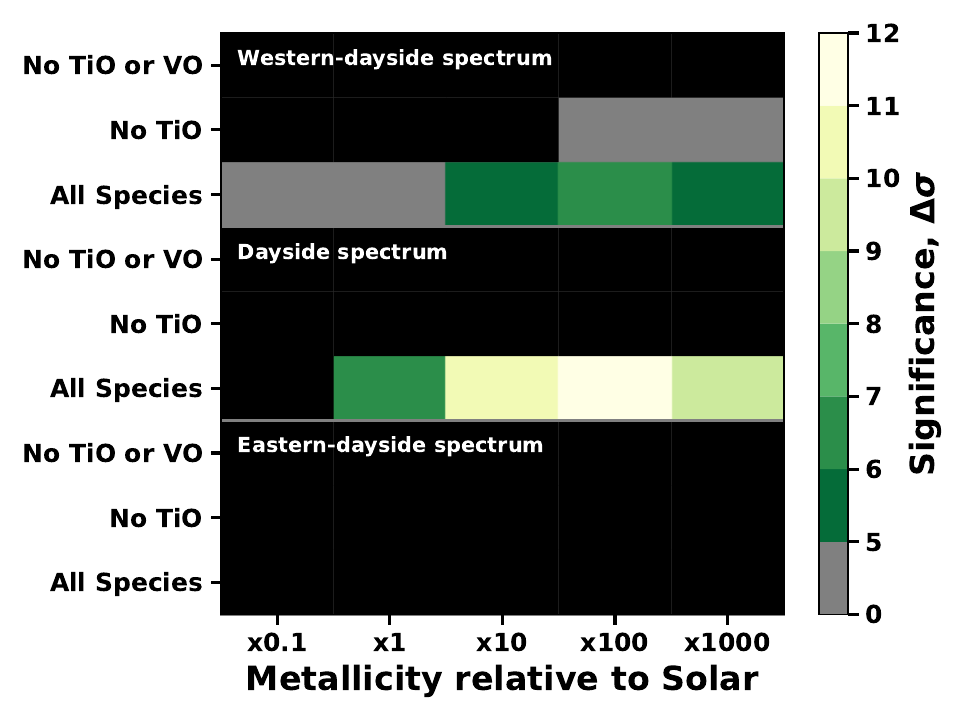}
    \caption{Results of the injection-recovery test on the `self-consistent' models with the original temperature-profile (T.1 in Table \ref{tab:models}). A black square indicates that the highest likelihood peak was more than $\SI{50}{\kilo\meter\per\second}$ in $V_{sys}$ or $K_p$ from the expected location of the injected orbit. Coloured squares indicate the significance ($\Delta\sigma$) at which the injected spectrum is recovered with grey squares indicating the models recovered at $\Delta\sigma < 5$.}
    \label{fig:recoveryltt}
\end{figure}

\begin{figure}
	\includegraphics[width=\columnwidth]{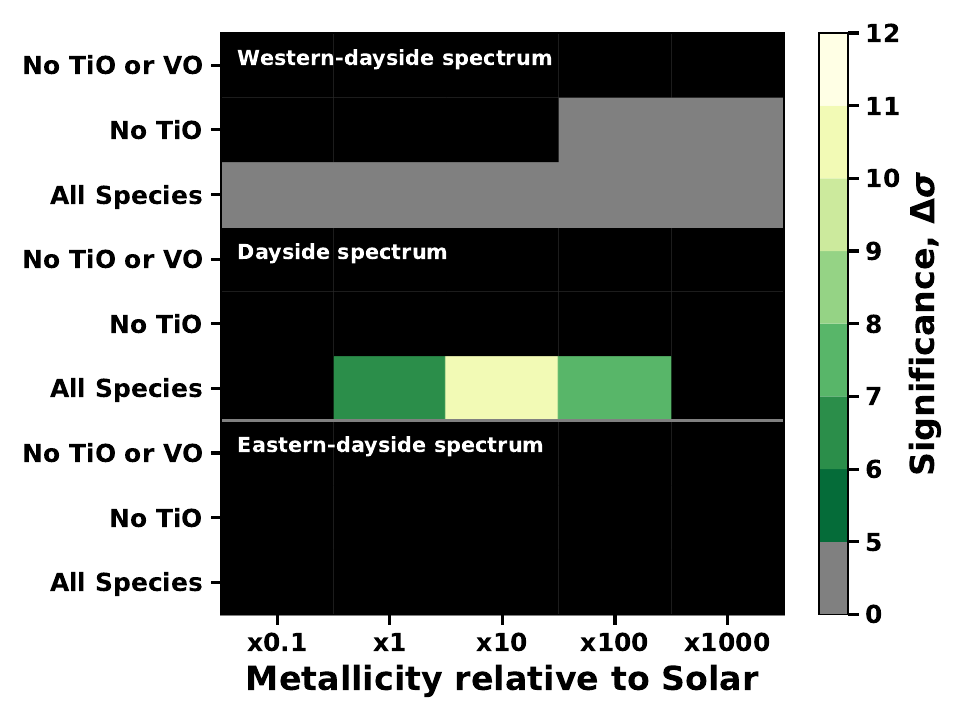}
    \caption{Same as Figure \ref{fig:recoveryltt} but for models created with the modified isothermal temperature-profile (T.2 in Table \ref{tab:models}).}
    \label{fig:recoveryltt_isop}
\end{figure}

In Figures \ref{fig:recoveryltt} and \ref{fig:recoveryltt_isop}, the significance at which a given model is recovered is indicated by the colour of the associated square where black indicates that the model was not recovered. We consider $\Delta\sigma > 5$ as the threshold for a confident detection. For the models with the original temperature-profile (T.1 in Table \ref{tab:models}), only the high metallicity dayside and western-dayside models containing all species are detectable. With the modified temperature-profile (T.2 in Table \ref{tab:models}), only the intermediate metallicity dayside models are detectable. These correspond to models with higher albedo and a large number of deep spectral lines. An interesting point to note is that the recovered significance matches with predictions made with synthetic observations used to design these observations (see Section \ref{sec:observations}).

\begin{table}
	\begin{center}
	\caption{Phase range split used in this analysis.}
	\label{tab:phasesplit}
	\begin{tabular}{rc} 
		\hline
		 & Phase (p) range \\
		\hline
        \hline
        Eastern-dayside & $p<0.417$ \quad ($150^\text{o}$) \\
        Dayside & $p>0.417$ and $p<0.583^\dagger$ \\
        Western-dayside & $p>0.583$ \quad ($210^\text{o}$) \\
		\hline
	\end{tabular}
    \end{center}
    $^\dagger$ Excluding secondary eclipse.
\end{table}

\subsubsection{Sensitivity to the `single-species' spectral models}
\label{sec:gridrecovery}

The `single-species' models are introduced in Section \ref{sec:searchgridtemplates} with each model containing one species (TiO, VO, Fe, MgH, FeH, H$_2$O or AlH). The models for each species are split into two sets, one that matches the albedo measured by \citet{Coulombe2025_dataset} in pre-eclipse JWST NIRISS/SOSS observations (C.1 in Table \ref{tab:models}) and another matching the post-eclipse (C.2 in Table \ref{tab:models}). For each species and each set, the models form a grid in which the VMR and the height of the cloud deck are varied. 

Since the two sets of models, C.1 and C.2 in Table \ref{tab:models}, are associated with different orbital phases, we once again inject these models into the observations of the corresponding orbital phase. The data is reduced as described in Section \ref{sec:datareduction} and the reduced observations split up into pre- and post-eclipse. The likelihood analysis in Section \ref{sec:recoverplanet} is performed on these two sets separately using the injected spectrum as the model in the likelihood analysis ($g$ in equation \ref{equ:likelihood}). We note that as discussed in Section \ref{sec:falsepositive}, some of the iron models have spectral lines much broader than the resolution element. For these we downsample the velocity resolution assuming the average broadening; $\SI{20}{\kilo\meter\per\second}$. 

\begin{figure}
	\includegraphics[width=\columnwidth]{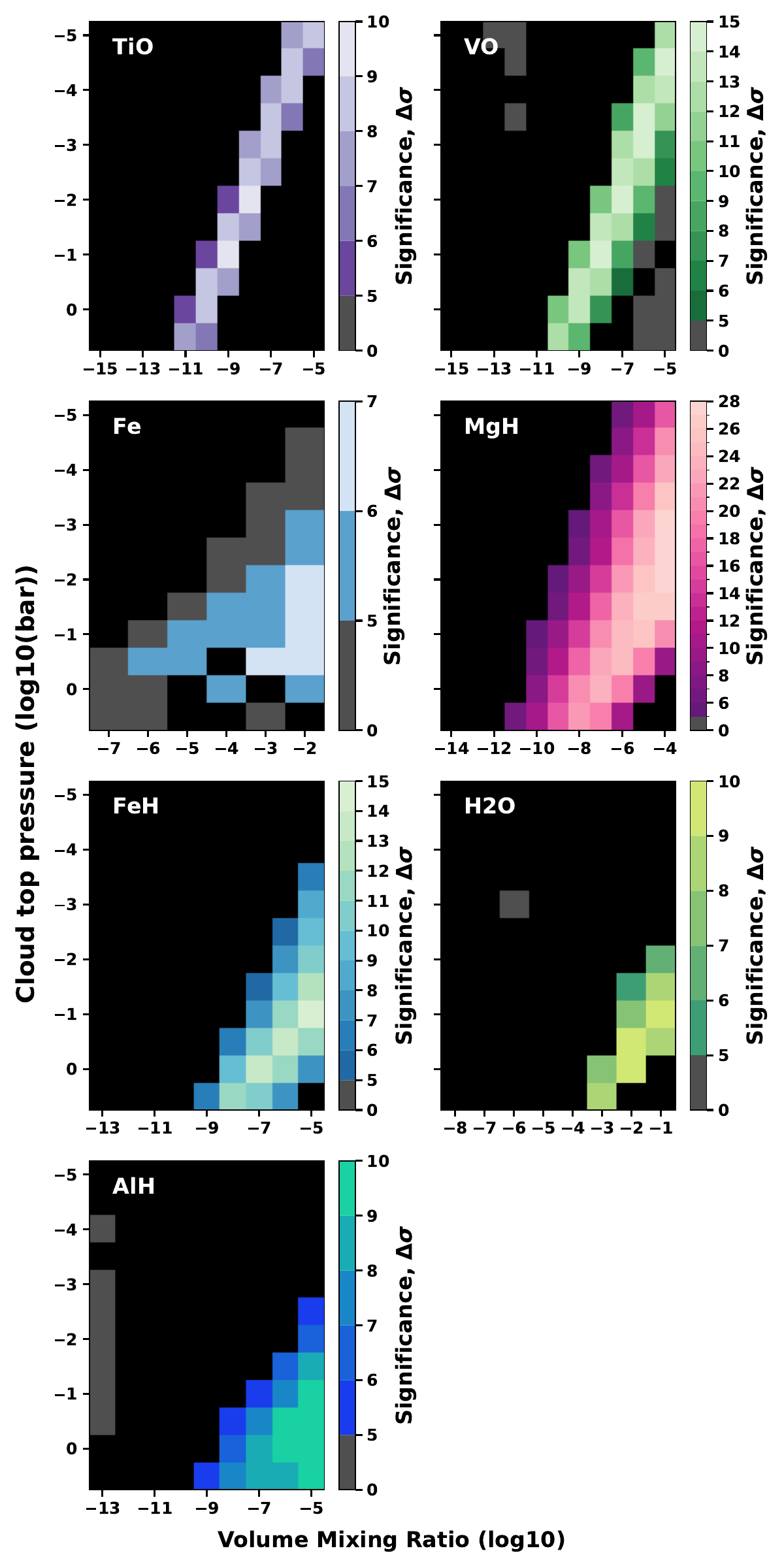}
    \caption{Significance, $\Delta\sigma$, at which the model is recovered in the injection-recovery test on the `single-species' models for the post-eclipse observations. A black square indicates that the model was not recovered in the injection-recovery tests. Coloured squares indicate the significance at which the injected spectrum is recovered with grey squares indicating the models recovered at $\Delta\sigma < 5$.}
    \label{fig:recoverygridpost}
\end{figure}

\begin{figure}
	\includegraphics[width=\columnwidth]{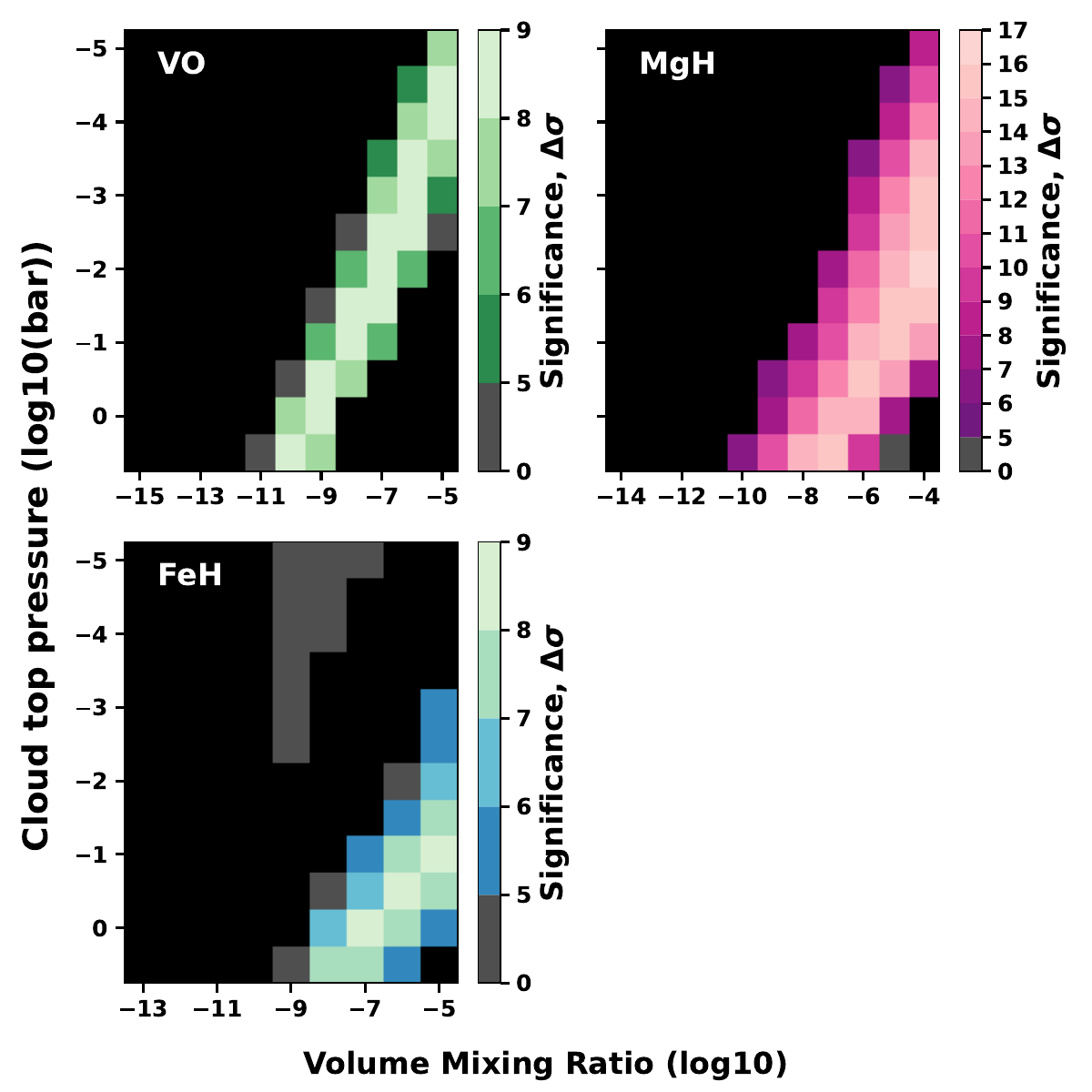}
    \caption{As Figure \ref{fig:recoverygridpost} but for the pre-eclipse observations.}
    \label{fig:recoverygridpre}
\end{figure}

Figure \ref{fig:recoverygridpost} show the significance, $\Delta\sigma$, with which each injected model is recovered for the post-eclipse data and Figure \ref{fig:recoverygridpre} shows the same for the pre-eclipse. Again, a black square in the plots indicates the model was not recovered. In post-eclipse we are able to recover some of the TiO, VO, Fe, MgH, FeH and H$_2$O models whereas in pre-eclipse we are only to recover some VO, MgH and FeH models. This is because even though we have more pre-eclipse data, the planet's eastern hemisphere has a lower albedo \citep{Coulombe2025} making its spectrum fainter and harder to characterise. The models with lower altitude cloud decks and higher abundances typically have deeper lines and are therefore detected at higher significance. This is not true of all models as in some cases the spectral lines start strongly saturating. When this happens, the saturation lowers the continuum of the spectrum which decreases the difference in contrast between the continuum and the minimum of the spectral line. This further lowers the signal-to-noise of the spectral lines embedded in the data and therefore the signal-to-noise at which the planet’s spectrum can be recovered. Additionally, the saturating spectral lines tend to become much broader which cause the continuum normalisation to absorb some of the line. This lowers the contrast between the line and continuum and thus lowers the recovered signal-to-noise of the spectrum. The combination of these effects makes some of the models with the highest abundances and deepest cloud decks undetectable with these data. 

Section \ref{sec:EUvmr} introduces a method of identifying models that would be recovered at similar significance by comparing their EU$_{\text{VMRs}}$. For the `single-species' models, the EU$_{\text{VMR}}$ is equivalent to the VMR of the species. Therefore these injection tests represent the range of EU$_{\text{VMRs}}$ that are detectable for each species, for a given cloud deck height in these data. In Table \ref{tab:constraints}, we summarise the EU$_{\text{VMRs}}$ of each species that can be ruled out as a result of this non-detection assuming the cloud deck is at the altitude predicted by the JWST NIRISS/SOSS data. With the exception of TiO in the post-eclipse data (see Section \ref{sec:tiodepletion} for further discussion) these restrictions are consistent with equilibrium chemistry. Due to the observed wavelength range, we are not sensitive to CH$_4$, CO and CO$_2$ hence we are unable to restrict the EU$_{\text{VMRs}}$ of these species.

Comparing these restrictions with the literature, we see that the pre-eclipse restriction on FeH is consistent with the work of \citet{Edwards2023a} and \citet{Zhou2025} which constrained the VMR${_\text{FeH}}\sim10^{-8}$ in the terminator atmosphere. Additionally, the restrictions on TiO and VO are more consistent with the constraints imposed by \citet{Edwards2023a} of VMR${_\text{TiO}}<10^{-8}$ and VMR${_\text{VO}}<10^{-9}$ for the terminator atmosphere over those imposed by \citet{Reyes2025} which constrained the metallicity to greater than $180\times$ solar corresponding to VMR${_\text{TiO}}\gtrsim10^{-7}$ and VMR${_\text{VO}}\gtrsim10^{-8}$. The discrepancy between the constraints imposed by these two works is due to the clouds of \lttb\ which flatten the transit spectrum. \citet{Reyes2025} do not include clouds in their model so the flattening of the spectrum is instead caused by increasing the TiO and VO abundances. The clouds in \citet{Edwards2023a} are lower in the atmosphere than assumed in Table \ref{tab:constraints}, hence their lower constraint on the VMRs of TiO and VO.

\begin{table*}[]
\begin{center}
\caption{Restrictions placed on the EU$_{\text{VMR}}$ of different species by these data.}
\label{tab:constraints}
\begin{tabular}{rcccccccccccccc}
\hline
                   & \multicolumn{7}{c}{Pre-eclipse (Cloud Top Pressure $10^{-1.5}$ bar)}    & \multicolumn{7}{c}{Post-eclipse (Cloud Top Pressure $10^{-4}$ bar)}  \\                          
Species            & TiO &     VO $\ddagger$    & Fe &     MgH    &     FeH $\dagger$    & H$_2$O & AlH          &    TiO $\ddagger$    &    VO $\ddagger$     & Fe &    MgH     & FeH & H$_2$O & AlH  \\
\hline
\hline
EU$_{\text{VMR}}$  & (...) & $<10^{-8}$ & (...) & $<10^{-7}$ & $<10^{-6}$  & (...)  & (...) & $<10^{-7}$ & $<10^{-6}$ & (...) & $<10^{-7}$ & (...) & (...) & (...) \\
\hline
\end{tabular}
\end{center}
\end{table*}

The restrictions on the EU$_{\text{VMR}}$ imposed by these injections come with a caveat: since the model injected and the model used in the likelihood analysis are the same, they represent a `best case scenario’. This is because the real spectrum will be slightly different to the model and also contain the spectral lines of other species which may interfere with the lines of the chosen species. A more comprehensive test would have been to inject a full model containing all species, which is then recovered with the `single-species' models. However, since we are not assuming equilibrium chemistry, this leads to an infeasibly large number of models containing all species at varying abundances and thus it was not possible to perform these injection tests.

\subsubsection{Determining the sensitivity to new spectral models}
\label{sec:recoverycomaprison}

As discussed in Section \ref{sec:EUvmr}, models with similar numbers and depths of spectral lines are expected to be recovered at similar significances. The number and depth of spectral lines is similar for models with the same EU$_{\text{VMRs}}$. Further, the `single-species' injection tests reveal the range of EU$_{\text{VMRs}}$ for each species that would be detectable with this data as a function of the altitude of the top of the cloud deck. Therefore, by computing the EU$_{\text{VMR}}$ of a more complex model, and comparing it to the results of the `single-species' injection test, we can determine if that model should have been detected with these data without the need for further injection tests. Additionally, as the `single-species' models did not result in a detection, these more complex models can be ruled out. 

\begin{figure}
    \centering
	\includegraphics[width=0.9\columnwidth]{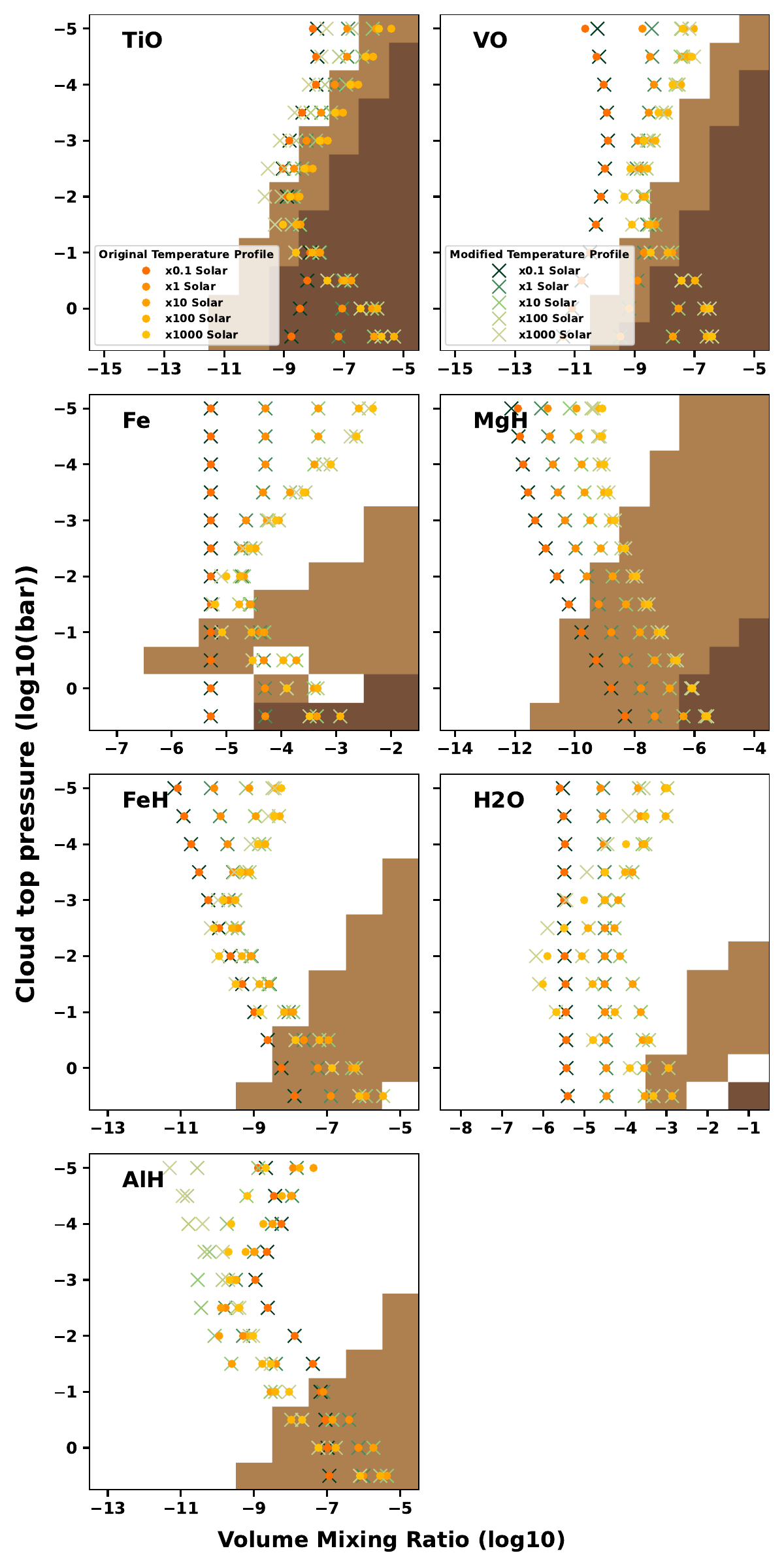}
    \caption{Restrictions placed on the VMR as a function of cloud deck altitude for the post-eclipse observations. The injection recovery tests presented here rule out the models coloured in light brown. Additionally, we highlight the models in dark brown which, due to the saturation of a large number of the spectral lines present, have an average contrast ratio more than three sigma below the contrast of \lttb\ at these wavelengths measured by JWST NIRISS/SOSS data. To compare these restrictions to more complex models, we also plot the EU$_{\text{VMRs}}$ for metallicities ranging from $0.1\times$ solar to $1000\times$ solar as a function of the altitude of the cloud deck for both the original (T.1, circles) and modified (T.2, crosses) temperature-profiles for the western-dayside.}
    \label{fig:recoverygridpost_companion}
\end{figure}

\begin{figure}
    \centering
	\includegraphics[width=0.9\columnwidth]{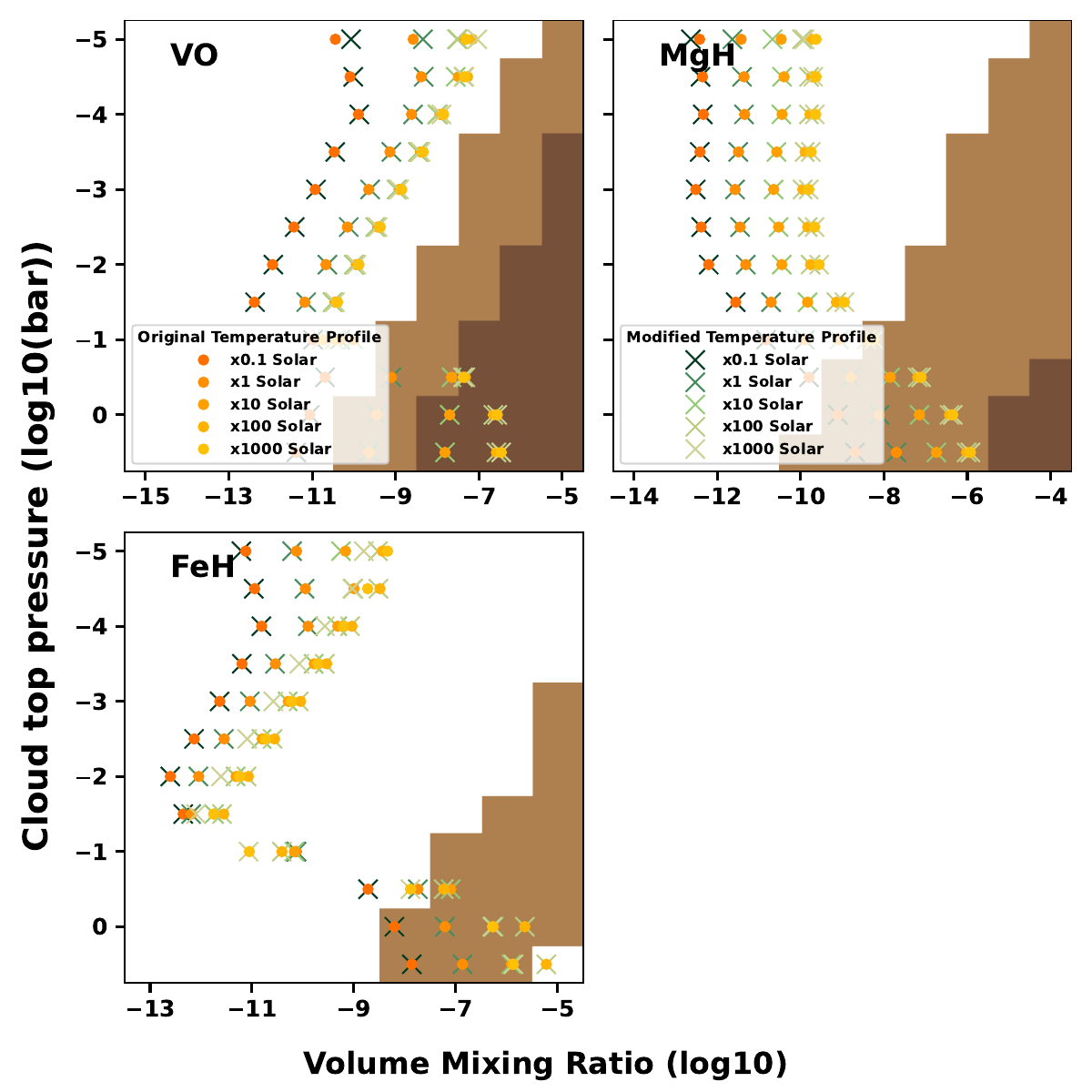}
    \caption{As Figure \ref{fig:recoverygridpost_companion} but for the pre-eclipse observations and eastern-dayside EU$_{\text{VMRs}}$.}
    \label{fig:recoverygridpre_companion}
\end{figure}

We demonstrate this method for ruling out new models in Figures \ref{fig:recoverygridpost_companion} and \ref{fig:recoverygridpre_companion}. In these plots we once again show the results of the `single-species' injection test but with different colour coding. Here, all the models recovered with $\Delta\sigma>5$ are highlighted in light brown. Additionally, highlighted in dark brown are models for which the saturation of spectral lines has reduced the average contrast ratio of the model to at least three sigma below the contrast of \lttb\ measured by JWST NIRISS/SOSS. Together, these two regions highlight the models ruled out but these and previous data. Next, we compute the EU$_{\text{VMRs}}$ as a function of the height of the cloud deck for the full `self-consistent' chemical equilibrium model (Ch.1 in Table \ref{tab:models}) for the eastern-dayside and western-dayside. These profiles cover a range of metallicities from $0.1\times$ solar to $1000\times$ solar and two versions of the temperature-profile (T.1 and T.2 in Table \ref{tab:models}). 

The EU$_{\text{VMRs}}$ as a function of the height of the cloud deck for western-dayside (approximately post-eclipse) are plotted on Figure \ref{fig:recoverygridpost_companion} and those for the eastern-dayside (approximately pre-eclipse) on Figure \ref{fig:recoverygridpre_companion}. This shows, if we were to vary the altitude of the cloud in the `self-consistent' models, whether the resulting spectrum would have been detectable in these data. 

The predictions made using the EU$_{\text{VMRs}}$ match closely to the results of the `self-consistent' model injections tests shown in Figures \ref{fig:recoveryltt} and \ref{fig:recoveryltt_isop}. For example, the EU$_{\text{VMRs}}$ of each species in the pre-eclipse data shown in Figure \ref{fig:recoverygridpre_companion} do not fall within the detectable region at a cloud top pressure of $10^{-1.5}$ bar. This is consistent with none of the eastern-dayside `self-consistent' models being recovered in the injection tests. Likewise for the post-eclipse data, the EU$_{\text{VMRs}}$ of TiO predicts that the $1\times$ to $1000\times$ solar models for the original temperature-profile (T.1 in Table \ref{tab:models}) and the $1\times$ and $10\times$ solar models for the modified temperature-profile (T.2 in Table \ref{tab:models}) will be detectable for a cloud top pressure of $10^{-4}$ bar. This matches with the injection tests for the dayside models, which have a similar albedo to the post-eclipse models. The exception is the dayside $100\times$ solar, modified temperature-profile model which is detected in the injection test but is not predicted to be. Since the `self-consistent' models contain the multiple species, they will be recovered at higher significance than predicted using the EU$_{\text{VMR}}$ of a single species and this model has an EU$_{\text{VMR}}$ for TiO close to the detection limit. The predictions do not match as well with the injection tests for the western-dayside models where the injection tests did not recover, as predicted, the $1\times$ solar, original temperature-profile model or any of the modified temperature-profile models. This discrepancy is likely caused by the difference in the phase range of the injection tests. The injection tests used to predict the detectability in Figures \ref{fig:recoverygridpost_companion} and \ref{fig:recoverygridpre_companion} are for the entire pre- and post-eclipse data whereas the `self-consistent' model injections split the data up into eastern-dayside, dayside and western-dayside. The detection strengths of the `self-consistent' models do however correlate with the EU$_{\text{VMR}}$ since models with higher EU$_{\text{VMR}}$ are detected with greater significance, $\Delta\sigma$. This highlights the usefulness of this method of comparison. 

\section{Discussion}
\label{sec:discussion}

\subsection{On the non-detection of TiO}
\label{sec:tiodepletion}

Previous observations have suggested that TiO might be depleted in the atmosphere of \lttb. As discussed in Section \ref{sec:nondetection}, we do not detect the reflected light spectrum of \lttb\ in these data. However, as shown in Section \ref{sec:lttrecovery} we should have been sensitive to spectral models based on data from previous observations, assuming high metallicity with TiO at its equilibrium abundance, in the post-eclipse data. Therefore, this non-detection might indicate a depletion of TiO in the western-hemisphere of \lttb. To determine if our non-detection is a result of TiO depletion, we consider the following alternative possibilities: 1) TiO is present at its equilibrium abundance but the spectral lines are muted by a high-altitude cloud deck, 2) the temperature-profile is different to what was previously measured and TiO is not expected in significant quantities, and 3) the combination of these two alternatives. To explore these alternatives, we compare the EU$_{\text{VMRs}}$ of the `self-consistent' western-dayside models with the results of `single-species' TiO injection test shown in the top left panel of Figure \ref{fig:recoverygridpost_companion}. 

To explore whether TiO could still be present at its equilibrium abundance but hidden by a high-altitude cloud deck, we study the TiO EU$_{\text{VMRs}}$ for the western-dayside's original temperature-profile (T.1 in Table \ref{tab:models}). In the top left panel of Figure \ref{fig:recoverygridpost_companion} the TiO EU$_{\text{VMRs}}$ for the original temperature-profile are shown as circles. The high metallicity versions of these EU$_{\text{VMRs}}$ fall within either the light or dark brown ruled out regions. This implies that, if \lttb's atmosphere is high metallicity and the western dayside follows the original T.1 temperature-profile, then the cloud deck would need to be above $10^{-5}$ bar for these data to result in a non-detection of TiO assuming it is present at its equilibrium abundance. To understand if changing the temperature-profile changes the expected TiO abundance significantly enough to change this result, we also study the TiO EU$_{\text{VMRs}}$ for the modified temperature-profile (T.2 in Table \ref{tab:models}; shown as crosses in Figure \ref{fig:recoverygridpost_companion}). The modified temperature-profile lowers the EU$_{\text{VMRs}}$ in the highest and lowest metallicity models. This results in cloud decks above $10^{-2}$ bar becoming effective at shielding the presence of TiO at its equilibrium abundance for these metallicities. However, for intermediate ($10\times$ solar) metallicities, the cloud deck would once again need to be over $10^{-5}$ bar to have the same effect. 

Ultimately, with the uncertainty in the upper temperature-profile and the metallicity, it is not possible to say for certain if our non-detection is a result of TiO depletion in the western atmosphere of \lttb. However, since the non-detection of TiO would require extreme cloud altitudes for more than half of our high metallicity models, we argue that depletion is currently the more likely reason for our non-detection. This conclusion comes with a caveat: we assume our TiO line list is accurate enough that the model lines would correlate well with any real TiO lines present in the planet's spectrum. We use the same line list as in \citet{Prinoth2022}, \citet{Prinoth2023} and \citet{Prinoth2025}. In the former two works, this line list was able to detect the presence of TiO on WASP-189\,b and in the latter a non-detection of TiO on WASP-121\,b is presented. For WASP-189\,b, the signal of TiO recovered is five to six times weaker than that of an injected isothermal model which is attributed to inaccuracies in the line lists. Similarly, for WASP-121\,b a detection of Ti I is in tension with the non-detection of TiO which may indicate that inaccuracies in the line list are responsible for the non-detection as opposed to a lack of TiO. These works would imply that our non-detection of TiO is largely due to inaccuracies in the line lists. However, subsequent low resolution observations WASP-121\,b have indicated a strong depletion of TiO \citep{Pelletier2025, Saha2025b}. We argue that, ultimately, it is difficult to quantify how well the line list is performing without knowing for certain what the true abundance of TiO is on the exoplanets being studied which, as we have highlighted in this work, is difficult to determine.

\subsection{Constraints the cloud properties}
\label{sec:othercosntraints}

\lttb\ is a cloudy world. Studies of the terminator atmosphere have favoured a low cloud deck at approximately $1$ bar \citep{Edwards2023a, Zhou2025}. Since the dayside of the planet is hotter than the terminator, we would expect to find a higher cloud deck on the dayside. However, the altitude of the dayside cloud deck is not well constrained \citep{Coulombe2025}. As discussed in Section \ref{sec:introduction}, understanding this world and its place in the desert requires detailed information on its atmosphere. Current studies are hindered by the degeneracy between the altitude of the cloud deck, and the metallicity of this atmosphere. These data allows us to improve restrictions on the altitude of the cloud deck, thus partly breaking this degeneracy. 

The restrictions we place on the altitude of the cloud deck are informed by the non-detection of MgH in both the pre- and post- eclipse data (see Section \ref{sec:nondetection}). The `single-species' injection test for MgH, shown in Figures \ref{fig:recoverygridpost} and \ref{fig:recoverygridpre}, reveals the range of EU$_{\text{VMRs}}$ for MgH that can be ruled out as a result of this non-detection. We can then compare this with the EU$_{\text{VMRs}}$ for MgH predicted from equilibrium chemistry models to determine at what altitudes the cloud deck obscures the MgH spectral features enough to prevent detection. Figures \ref{fig:recoverygridpost_companion} and \ref{fig:recoverygridpre_companion} show EU$_{\text{VMRs}}$ for MgH predicted from equilibrium chemistry for two different temperature-profiles (T.1 and T.2 in Table \ref{tab:models}) along with the parameter space ruled out by this non-detection and previous data. Comparing the EU$_{\text{VMRs}}$ for the two temperature-profiles, it is clear that the MgH abundance is not very sensitive to temperature. This makes it ideal for investigating the altitude of the cloud deck. Assuming, as suggested by previous observations, that this planet has a high metallicity we compare the EU$_{\text{VMRs}}$ for the $10\times$, $100\times$ and $1000\times$ solar models. Figure \ref{fig:recoverygridpost_companion} indicates that our post-eclipse data should have been sensitive to MgH at the EU$_{\text{VMRs}}$ predicted for the high metallicity models when the top of the cloud deck is at pressures greater than $10^{-2}$ bar. Thus \lttb's cloud deck must be at pressures less than $10^{-2}$ bar in the western hemisphere. Likewise, Figure \ref{fig:recoverygridpre} indicates that our pre-eclipse data should have been sensitive to MgH if the cloud deck was at pressures greater than $10^{-0.5}$ bar, hence the top of the cloud deck in the eastern-hemisphere must be at lower pressures.

\subsection{A pathway to the detection of reflected light at high spectral resolution}
\label{sec:pathway}

These data were taken with ESPRESSO in 4-UT mode with just under 9 hours spent on pre-eclipse and 5 hours and 40 minutes on post-eclipse. These observations have an average S/N $\approx 160$ at $\lambda = \SI{530}{\nano\meter}$. We note that \ltt\ (V = 9.79 mag) is on the fainter end of stars that can be characterised at high resolution hence why we used the greater collecting area of ESPRESSO's 4-UT mode for these observations. Unfortunately, despite \lttb's high albedo, we do not detect its high-resolution reflection spectrum which implies this planet lacks the large number of deep spectral lines required for this technique to work. 

This highlights that the broadband albedo is not necessarily a good indication of detectability with the HRCCS technique as this does not necessarily imply the spectrum has enough deep lines to be detected. Thus we suggest the best targets to characterise with this technique using current instrumentation have the following properties; 1) A mid-to-high albedo as this increases the signal-to-noise of the planets spectrum, 2) A low altitude cloud deck, and 3) High metallicity. In addition to these three points it would also be more favourable to have 1) A brighter host star than \ltt, 2) A longer orbital period to reduce the rotational broadening of the reflected stellar spectrum so that the stellar lines could be used in the cross-correlation templates, and 3) A planet in the transiting configuration because these tend to have much more precise ephemerides. 

It is not long before the ELT comes online. Its ANDES instrument will be able to take observations similar to ESPRESSO but at $R=100,000$, over a larger wavelength range $0.4-\SI{1.8}{\micro\meter}$, and on a larger telescope \citep{Palle2023, Marconi2024}. The ANDES ETC \footnote{\url{https://andes.inaf.it/instrument/exposure-time-calculator/}} indicates that ANDES will reach an $S/N \approx 740$ at $\lambda = \SI{530}{\nano\meter}$ in $\SI{200}{\second}$ assuming a G8V star with a V-band magnitude of 9.79 at an airmass of 1.3. When compared to these ESPRESSO observations which reach $S/N \approx 160$ at $\lambda = \SI{530}{\nano\meter}$ in $\SI{200}{\second}$. This is a 4.6 times improvement in the signal-to-noise as a function of time which is a result of the larger primary mirror and better instrumental throughput. This vast improvement in the signal-to-noise and extended wavelength range will open up the characterisation of \lttb\ as other atmospheric species are predicted to become characterisable with the improvement in signal-to-noise using a similar amount of observing time to that already obtained. Additionally, this will open up the characterisation of even more worlds in reflected light from giant planets to potentially some rocky worlds \citep[see e.g.][]{Palle2023}. While characterising rocky worlds will be challenging, simulations of similar observations using HARMONI, which will also be on the ELT but has a lower spectral resolution, have shown that such worlds can be characterised \citep{Vaughan2024}. In all cases, the detectability of a planet's reflected light spectrum at high spectral resolution is dependent on it possessing a large number of deep spectral lines.

\section{Conclusions}
\label{sec:conclusions}

We use four half-nights of ESPRESSO 4-UT mode observations to characterise the reflected light spectrum of \lttb\ at high spectral resolution. From these we conclude the following:

   \begin{enumerate}
      \item In these data, we are unable to robustly detect the reflected light spectrum of \lttb\ despite its known high albedo. We would have expected to detect the presence of TiO in the dayside and western-dayside spectrum for a range of different metallicities and assumptions on the temperature-profile. Therefore this non-detection could imply a depletion of TiO compared with equilibrium chemistry. This depletion can be caused by the cold-trapping of TiO and is consistent with the lack of a thermal inversion \citep{Dragomir2020}. However, we have also shown that the abundance of TiO in chemical equilibrium and therefore its detectability is sensitive to the temperature in the upper atmosphere at $P<10\mu \text{bar}$. Therefore, it is not currently possible to determine if the non-detection of TiO is due to depletion given the uncertainty in its equilibrium abundance.  
      \item A non-detection of MgH in post-eclipse indicates that the clouds on the western-dayside are at pressures lower than $10^{-2}$ bar and a non-detection of MgH in pre-eclipse indicates the eastern-dayside clouds are at pressures lower than $10^{-0.5}$ bar. This is consistent with predictions of MgSiO$_3$ and  Mg$_2$SiO$_4$ clouds based on the mean JWST NIRISS/SOSS temperature-profile.
      \item The atmospheric characterisation presented in this work also highlighted a false positive detection. We determined that this was due to oversampling the $V_{sys}$-$K_p$ map, as in this case the majority of the Fe lines are broader than the instrumental resolution, which created correlated noise. Thus we highlight caution when interpreting HRCCS detections made with models containing broad spectral lines. 
      \item In addition to the atmospheric characterisation, we also review the orbit of this planet. By combining these data with previous observations we have computed an updated ephemeris. Our orbit fits favour a circular orbit however there is some indication of non-zero eccentricity. This requires more data for confirmation, but if confirmed it may be an indication of high eccentricity migration. 
      \item Finally, we discuss the future of high spectral resolution characterisation of exoplanets in reflected light with the VLT and the upcoming ELT. The unrivalled collecting area of the ELT will further the characterisation of \lttb\ and open the door to many other worlds. However, as we have highlighted on this work, even very reflective exoplanets may not be detectable at high spectral resolution if they lack deep spectral lines. As such, we advise caution when selecting targets for observations.
   \end{enumerate}

\begin{acknowledgements}

This work is based on observations collected at the European Organisation for Astronomical Research in the Southern Hemisphere under ESO programmes 0112.C-0367, 2103.C-5063, 0108.C-0161, 0102.C-0525, and 0102.C-0451.

SRV, JLB and LTP acknowledge funding from the European Research Council (ERC) under the European Union’s Horizon 2020 research and innovation program under grant agreement No. 805445.
J.S.J gratefully acknowledges support by FONDECYT grant 1240738 and from the ANID BASAL projects ACE210002 and FB210003.
H.Y. acknowledge funding from the European Research Council under the European Union’s Horizon 2020 research and innovation programme (grant agreement No 865624, GPRV).
JVS acknowledges support from the Poincaré fellowship of the Observatoire de la Côte d'Azur.
VP acknowledges funding by the French National Research Agency (ANR) project EXOWINDS (ANR-23-CE31-0001-01).
N.E.B acknowledges support from NASA’S Interdisciplinary Consortia for Astrobiology Research (NNH19ZDA001N-ICAR) under award number 19-ICAR19 2-0041.

The first author would also like to thank Paul Mollière for their discussion on atmospheric column mass and Lorena Acuna for their discussion on tidal heating.

\end{acknowledgements}

%
\bibliographystyle{aa} 
\bibliography{aa57240-25.bib} 
%

\begin{appendix} 

\section{Additional Figures and Tables}

\begin{table}[h]
\centering
\caption{Line lists used by PICASO.}
\begin{tabular}{cc}
    \hline
    Species & Reference \\
    \hline
    \hline
    H$_2$ &  \citet{HITRAN2016} \\ 
    AlH & \citet{GharibNezhad2021}  \\ 
    CO &  \citet{HITEMP2010,HITRAN2016,li15rovibrational} \\ 
    CO$_2$ &  \citet{HUANG2014reliable} \\ 
    CaH &  \citet{Yadin2012MgH, GharibNezhad2021}  \\ 
    CrH & \citet{Burrows02_CrH, GharibNezhad2021}  \\ 
    Fe &  \citet{Ryabchikova2015,oBrian1991Fe,Fuhr1988Fe,Bard1991Fe,Bard1994Fe} \\ 
    FeH &  \citet{Dulick2003FeH,Hargreaves2010FeH, GharibNezhad2021} \\ 
    H$_2$O &  \citet{Polyansky2018H2O, GharibNezhad2021} \\ 
    H$_2$S &  \citet{azzam16exomol} \\ 
    H3+ &  \citet{Mizus2017H3p} \\
    HCN &  \citet{Harris2006hcn,Barber2014HCN,hitran2020} \\ 
    K &  \citet{Ryabchikova2015,Allard2007AA,Allard2007EPJD,Allard2016,Allard2019} \\ 
    Li & \citet{Ryabchikova2015,Allard2007AA,Allard2007EPJD,Allard2016,Allard2019}  \\ 
    LiH & \citet{Coppola2011LiH, GharibNezhad2021}  \\ 
    MgH &  \citet{Yadin2012MgH,GharibNezhad2013MgH,GharibNezhad2021} \\ 
    Na &  \citet{Ryabchikova2015,Allard2007AA,Allard2007EPJD,Allard2016,Allard2019} \\ 
    O$_2$ & \citet{HITRAN2016} \\
    Rb & \citet{Ryabchikova2015,Allard2007AA,Allard2007EPJD,Allard2016,Allard2019} \\
    TiH &  \citet{GharibNezhad2021} \\
    VO &  \citet{McKemmish16,GharibNezhad2021} \\ 
    TiO &  \citet{McKemmish2019TiO,GharibNezhad2021} \\ 
    H2--H2 &  \citet{Saumon12,Lenzuni1991h2h2} \\ 
    H2--He &  \citet{Saumon12} \\ 
    \hline
    \end{tabular}
    \label{tab:opas}
\end{table}

\begin{figure*}
	\includegraphics[width=\textwidth]{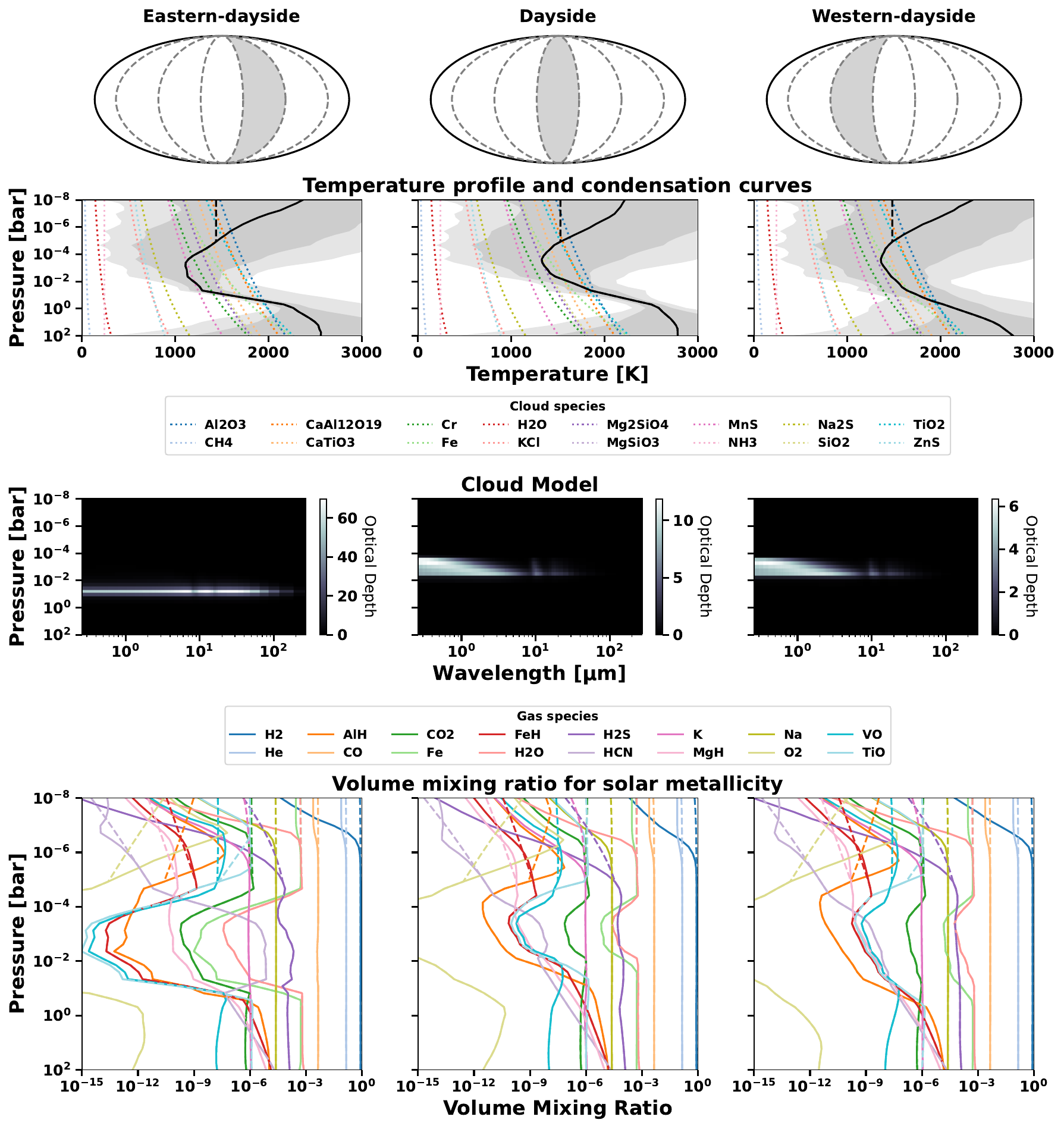}
    \caption{The inputs used to create the `self-consistent' spectra with PICASO for each atmospheric segment (top row). The second row shows the temperature-pressure profiles of each segment with the solid line indicating the original profile and the shaded regions the one and two sigma errors. The dash line indicates the modification to the profile for the isothermal models. The dotted lines are the condensation curves for different cloud species. Previous works have favoured the presence of Mg$_2$SiO$_4$ and MgSiO$_3$ clouds over those of other types \citep{Hoyer2023, Radica2024a, Coulombe2025, Radica2025} thus we only consider Mg$_2$SiO$_4$ and MgSiO$_3$ clouds in our models. The third row shows the optical depth per atmospheric layer of the modelled Mg$_2$SiO$_4$ and MgSiO$_3$ clouds as a function of wavelength. In the eastern-dayside model, the top of the cloud deck forms at $10^{-1.5}$ bar and for the other two segments it is at $10^{-4}$ bar. The last row shows the VMRs of all the species used in these models for $10\times$ solar metallicity. These assume equilibrium chemistry with the original temperature-profile (solid lines) and modified profile (dashed lines).}
    \label{fig:spectra_inputs}
\end{figure*}

\end{appendix}

\end{document}